\documentclass[journal, 10pt]{IEEEtran}

\usepackage{cite}
\usepackage{amsmath,amssymb,amsfonts,amsthm}
\usepackage{algorithm}
\usepackage{setspace}
\usepackage{algcompatible}
\usepackage{graphicx}
\usepackage{textcomp}
\usepackage{xcolor}
\usepackage{mathtools}
\usepackage{graphicx}
\usepackage{booktabs}
\usepackage{makecell}
\usepackage[font={small}]{caption}
\usepackage{subcaption}
\usepackage{cite}
\usepackage{breqn}
\usepackage{mathrsfs}
\usepackage{accents}
\usepackage{acronym}
\usepackage{soul}
\usepackage{bm}
\usepackage{url}
\usepackage{wrapfig}
\usepackage{todonotes}
\usepackage{verbatim}
\usepackage{tikz}
\usepackage{pdfpages}
\usepackage{multirow}

\DeclareMathOperator*{\argmin}{arg\,min}

\graphicspath{{./figures/}}
\acrodef{prop}[\textit{MIMORPH}]{MIMO Radio Platform for Heterogeneous wireless systems}
\acrodef{adc}[ADC]{Analog-to-Digital Converter}
\acrodef{aoa}[AoA]{Angle of Arrival}
\acrodef{aod}[AoD]{Angle of Departure}
\acrodef{ap}[AP]{Access Point}
\acrodef{bpsk}[BPSK]{Binary Phase-Shift Keying}
\acrodef{cdf}[CDF]{Cumulative Distribution Function}
\acrodef{cef}[CEF]{Channel Estimation Field}
\acrodef{cfo}[CFO]{Carrier Frequency Offset}
\acrodef{cir}[CIR]{Channel Impulse Response}
\acrodef{csi}[CSI]{Channel State Information}
\acrodef{cs}[CS]{Compressed Sensing}
\acrodef{cnn}[CNN]{Convolutional Neural Network}
\acrodef{dft}[DFT]{Discrete Fourier Transform}
\acrodef{dl}[DL]{Deep Learning}
\acrodef{dust}[DUST]{Deep Unfolding Sparse Transformer mode}
\acrodef{edmg}[EDMG]{Enhanced Directional Multi Gigabit}
\acrodef{ekf}[EKF]{Extended Kalman Filter}
\acrodef{elu}[ELU]{Exponential-Linear Unit}
\acrodef{fmcw}[FMCW]{Frequency-Modulated Continuous-Wave}
\acrodef{fov}[FOV]{Field-of-View}
\acrodef{ft}[FT]{Fourier Transform}
\acrodef{har}[HAR]{Human Activity Recognition}
\acrodef{if}[IF]{Intermediate Frequency}
\acrodef{ifs}[IFS]{Inter-Frame Spacing}
\acrodef{iht}[IHT]{Iterative Hard-Thresholding}
\acrodef{ista}[ISTA]{Iterative Shrinkage-Thresholding Algorithm}
\acrodef{isac}[ISAC]{Integrated Sensing And Communication}
\acrodef{jcs}[JCS]{Joint Communication and Sensing}
\acrodef{lasso}[LASSO]{Least Absolute Shrinkage and Selection Operator}
\acrodef{los}[LOS]{Line-of-Sight}
\acrodef{liht}[LIHT]{Learned Iterative Hard-Thresholding}
\acrodef{lista}[LISTA]{Learned Iterative Shrinkage-Thresholding Algorithm}
\acrodef{mae}[MAE]{Mean Absolute Error}
\acrodef{mse}[MSE]{Mean Squared Error}
\acrodef{md}[$\mu$D]{micro-Doppler}
\acrodef{mimo}[MIMO]{Multiple Input Multiple Output}
\acrodef{mmwave}[mmWave]{Millimeter-Wave}
\acrodef{MUSIC}[MUSIC]{MUlti SIgnal Classification}
\acrodef{nlos}[NLOS]{Non-Line-of-Sight}
\acrodef{nn}[NN]{Neural Network}
\acrodef{ofdm}[OFDM]{Orthogonal Frequency Division Multiplexing}
\acrodef{omp}[OMP]{Orthogonal Matching Pursuit}
\acrodef{phy}[PHY]{Physical Layer}
\acrodef{pov}[PoV]{Point-of-View}
\acrodef{psdu}[PSDU]{Physical layer Service Data Unit}
\acrodef{rf}[RF]{Radio Frequency}
\acrodef{relu}[ReLU]{Rectified Linear Unit}
\acrodef{rcs}[RCS]{Radar Cross-Section}
\acrodef{rss}[RSS]{Received Signal Strength}
\acrodef{rnn}[RNN]{Recurrent Neural Network}
\acrodef{rom}[ROM]{Read Only Memories}
\acrodef{rmse}[RMSE]{Root MSE}
\acrodef{sc}[SC]{Single Carrier}
\acrodef{sdr}[SDR]{Software Defined Radio}
\acrodef{siso}[SISO]{Single Input Single Output}
\acrodef{ssim}[SSIM]{Structural Similarity Index Metric}
\acrodef{sista}[SISTA]{Sequential ISTA}
\acrodef{slista}[SLISTA]{Sequential LISTA}
\acrodef{snr}[SNR]{Signal-to-Noise Ratio}
\acrodef{stft}[STFT]{Short Time Fourier Transform}
\acrodef{tf}[TF]{Time-Frequency}
\acrodef{toa}[ToA]{Time of Arrival}

\acrodef{cef}[CEF]{Channel Estimation Field}
\acrodef{mcs}[MCS]{Modulation and Coding Scheme}
\acrodef{stf}[STF]{Short Training Field}

\newcommand{\modelname}{STAR}
\newcommand{\eq}[1]{Eq.~\eqref{#1}}
\newcommand{\fig}[1]{Fig.~\ref{#1}}
\newcommand{\tab}[1]{Tab.~\ref{#1}}
\newcommand{\secref}[1]{Section~\ref{#1}}
\newcommand{\alg}[1]{Alg.~\ref{#1}}

\newcommand{\rev}[1]{{\color{blue}#1}}
\renewcommand{\rev}{}

\newcommand{\mytexttilde}{{\raise.17ex\hbox{$\scriptstyle\mathtt{\sim}$}}}

\begin{document}

\title{Attention-Refined Unrolling for Sparse \\ Sequential micro-Doppler Reconstruction}

\author{Riccardo Mazzieri$^{\ddag *}$,~\IEEEmembership{Graduate Student Member,~IEEE,}
        Jacopo Pegoraro$^{\ddag }$,~\IEEEmembership{Member,~IEEE,}
        and~Michele~Rossi$^{\dag \ddag}$,~\IEEEmembership{Senior~Member,~IEEE}
\thanks{$^{\ddag}$These authors are with the Department of Information Engineering at the University of Padova.
$^{\dag}$These authors are with the Department of Mathematics ``Tullio Levi-Civita'' at the University of Padova. $^{*}$Corresponding author email: \texttt{riccardo.mazzieri@phd.unipd.it}.

This work was partially supported by the European Union under the Italian National Recovery and Resilience Plan (NRRP) of NextGenerationEU, partnership on “Telecommunications of the Future” (PE0000001 - program “RESTART”).}}



\IEEEoverridecommandlockouts
\IEEEaftertitletext{\vspace{-1.5\baselineskip}}
\maketitle

\begin{abstract}
\rev{
The reconstruction of micro-Doppler signatures of human movements is a key enabler for fine-grained activity recognition wireless sensing. In \ac{jcs} systems, unlike in dedicated radar sensing systems, a suitable trade-off between sensing accuracy and communication overhead has to be attained. It follows that the micro-Doppler has to be reconstructed from \textit{incomplete} windows of channel estimates obtained from communication packets.   

Existing approaches exploit compressed sensing, but produce very poor reconstructions when only a few channel measurements are available, which is often the case with real communication patterns. In addition, the large number of iterations they need to converge hinders their use in real-time systems.

In this work, we propose and validate \modelname{}, a neural network that reconstructs micro-Doppler sequences of human movement even from highly incomplete channel measurements. \modelname{} is based upon a new architectural design that combines a \textit{single} unrolled iterative hard-thresholding layer with an attention mechanism, used at its output. This results in an interpretable and lightweight architecture that reaps the benefits of both model-based and data driven solutions.

\modelname{} is evaluated on a public \ac{jcs} dataset of $60$~GHz channel measurements of human activity traces. Experimental results show that it substantially outperforms state-of-the-art techniques in terms of the reconstructed micro-Doppler quality. Remarkably, \modelname{} enables human activity recognition with satisfactory accuracy even with $90$\% of missing channel measurements, for which existing techniques fail.}
\end{abstract} 


\IEEEpeerreviewmaketitle

\section{Introduction}
\label{sec:introduction}

\IEEEPARstart{N}{ext}-generation wireless networks are expected to gain the capability of sensing their surroundings via \ac{rf} signals, in addition to their primary communication functionality~\cite{wymeersch2021integration}. The vast number of applications of such context-aware networks spans domains like remote healthcare~\cite{shah2019rf}, safety~\cite{guo2023joint}, vehicle and crowd monitoring~\cite{kumari2017ieee}, and touchless human-computer interaction~\cite{regani2021mmwrite}, which all require real-time processing of a huge amount of raw sensing data. Moreover, advanced \ac{jcs} systems that analyze the movement of complex targets (e.g., humans) often involve computation-heavy \ac{dl} architectures~\cite{vandersmissen2018indoor, li2022unsupervised}. 
However, while research on \ac{jcs} is rapidly growing, there is an increasing concern that endowing communication systems with radar-like capabilities is bound to increase the network overhead and the channel occupation in time and frequency~\cite{wu2023green}. 

\begin{figure}
    \centering
    \includegraphics[width=\linewidth]{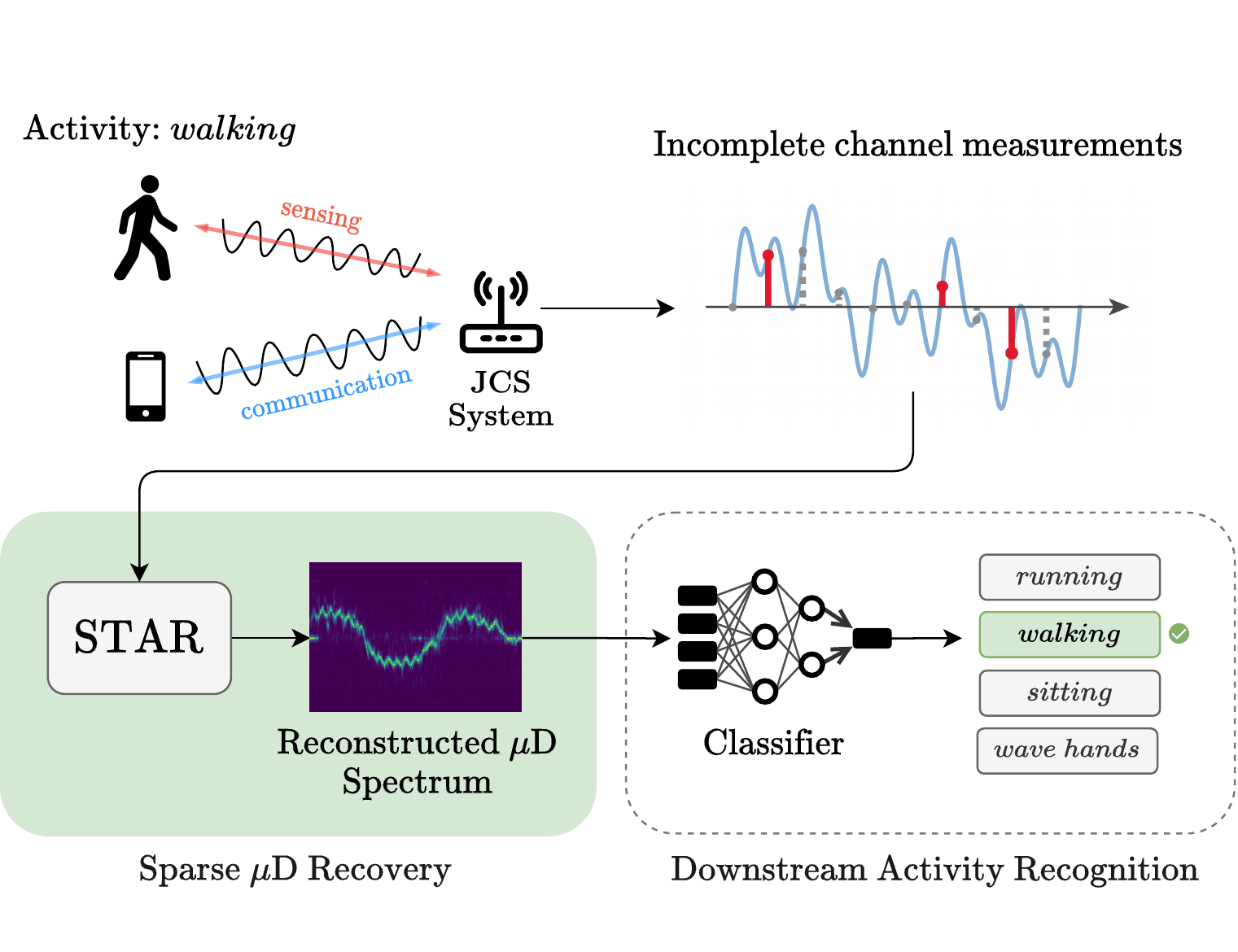}
    \caption{Processing chain for human activity recognition: \rev{our model (\modelname{})} is capable of recovering the \ac{md} spectrum from very few CIR measurements.}
    \label{fig:STAR_cover-diagram}
\end{figure}

In this work, we tackle the design of a {\it lightweight} and {\it ultra-low overhead} \ac{jcs} method for human movement analysis based on the \textit{sparse} reconstruction of the subject's \ac{md} signature. \ac{md} refers to the frequency modulation of the reflected radio signal caused by the motion of multiple target parts~\cite{chen2006micro}. \ac{md} is a widespread method for complex target recognition and motion analysis in radar~\cite{hanif2022micro} and, more recently, in \ac{jcs}~\cite{pegoraro2023rapid, pegoraro2022sparcs}. Its accurate computation usually requires regular and dense channel estimates, for the extraction of the Doppler spectrum. The main challenge of obtaining \ac{md} signatures in \ac{jcs} is that the overhead introduced by frequent channel estimations is excessive. Therefore, the Doppler estimation must be carried out relying on 
few and irregular channel estimates obtained from the communication packets that are naturally exchanged by the network terminals, which prevents the use of standard \ac{tf} analysis methods. 

To the best of our knowledge, only our previous work~\cite{pegoraro2022sparcs} has addressed this problem, devising a sparse reconstruction technique for the \ac{md} based on the \ac{iht} compressed sensing algorithm~\cite{eldar2012compressed}. However, this former approach has two main limitations: (i)~when faced with high 
amounts of missing measurements, e.g., $90$\%, it produces low-quality reconstructions of the \ac{md} that can lead to movement recognition errors, and (ii)~it takes many iterations to converge, which translate into a high computation cost and large processing delays. Note that both (i) and (ii) go against the critical requirements of an efficient \ac{jcs} system, which should operate with low overhead, performing very few channel measurements (i.e., avoiding the transmission of dummy packets for the sole purpose of sensing), and should guarantee fast reconstruction of the \ac{md}, enabling real-time sensing applications.

\rev{Additionally, classical compressed sensing approaches are not specifically tailored for \ac{md} spectrogram reconstruction and do not account for the temporal correlations of this type of data. 
Indeed, \ac{md} spectrograms exhibit specific temporal features, which may vary across different application domains. For example, in the case of human sensing, \ac{md} spectrograms are characterized by specific energy patterns, which might heavily differ from those of drones, cars, or other target types. Therefore, the key technical challenge tackled by this work is to design a system capable of improving upon classical compressed sensing methodologies, by leveraging domain-specific features of the signals to improve the quality of the reconstructions.
To do so, while meeting the requirements of a low-overhead \ac{jcs} system, we propose and validate Single Thresholding with Attention Refinement (\modelname{}). \modelname{} is an interpretable \ac{nn} architecture that accurately reconstructs \ac{md} signatures from highly incomplete time-domain channel measurements, with a computational complexity comparable to that of \textit{a single} \ac{iht} iteration.}

 
The considered processing chain for human activity recognition via wireless sensing is illustrated in \fig{fig:STAR_cover-diagram}: A wireless device samples the channel at irregular intervals, i.e., when data packets are transmitted over the wireless medium. This leads to the collection of 
incomplete vectors of channel measurements. 
\rev{Those are then translated by \modelname{} into the corresponding \ac{md} spectrum, which succeeds in this task even in the presence of a high percentage of missing measurements.} A classifier is finally used to assess the user activity.

 \modelname{} solves the drawbacks of prior techniques~\cite{pegoraro2022sparcs} by effectively combining: (1)~a model-based learning block, that obtains a candidate reconstruction by \textit{unrolling} a single \ac{iht} iteration into a \ac{nn} layer, (2)~a dot product attention layer with no learnable parameters, which exploits the sequential nature of the \ac{md} to learn its (temporal) correlation properties and to provide context features, and (3)~a solution refinement block, which improves the unrolled solution from step (1) using the context information from step (2), in an interpretable fashion, producing the final (refined) \ac{md} spectrum at its output. 

A thorough evaluation of the proposed architecture on experimental data is carried out, using the publicly available DISC dataset containing IEEE~802.11ay \ac{cir} measurements at $60$~GHz~\cite{pegoraro2022disc}. We focus on the task of reconstructing \ac{md} signatures of human movement for different activities, computing the reconstructed \ac{md} error with respect to the ground truth, and the resulting activity classification accuracy of a \ac{nn} classifier. \rev{The considered activities are walking, running, waving hands, and sitting down/standing up~\cite{pegoraro2022disc}.} \modelname{} shows superior performance to both the original \ac{iht} and state-of-the-art \ac{nn} models from the literature, providing accurate reconstructions even when only $10$\% of the input channel measurements are available.

\modelname{} paves the way for the utilization of lightweight interpretable \ac{nn} models to drastically lower the overhead of \ac{md} reconstruction in \ac{jcs}. \rev{Note that human activity recognition is just one of the possible applications of \modelname{}, which may be trained on any application involving \ac{md} recognition, e.g., drone and vehicle classification~\cite{hanif2022micro, gerard2021micro} and gait analysis~\cite{seifert2019toward}, among others.} 

The main contributions of this work are:
\begin{enumerate}
    \item We develop and validate \modelname{}, the first \ac{nn} model to reconstruct \ac{md} spectrograms from (very) 
    few
    time-domain estimates of the \ac{jcs} channel.
    \item \modelname{} features a new way of exploiting the sequential nature of the \ac{md}, by refining an unrolled \ac{iht} solution with context information extracted through an attention mechanism. The resulting \ac{md} spectrum quality is excellent, even with over $90\%$ missing channel measurements. 
    \item \modelname{} is lightweight, having the computational complexity of \textit{just one} \ac{iht} iteration. This makes it amenable to real-time operation in \ac{jcs} systems.
    \item We test \modelname{} on a publicly available \ac{jcs} dataset~\cite{pegoraro2022disc}, on the \ac{md} of human activities.
    When the \ac{md} signatures reconstructed from $90$\%-incomplete measurements are used for the classification of challenging activities, existing approaches completely fail, yielding \mbox{F$1$-scores} that approach zero. Conversely, \modelname{} provides \mbox{F$1$-scores} in the range $0.5-0.8$, showing a huge performance gain. 
\end{enumerate}
The \rev{paper} is organized as follows. In \secref{sec:rel-work} we discuss the related work, while \secref{sec:backg} introduces the necessary background on compressed sensing and deep unrolling. \secref{sec:sparse-md} presents our \ac{cir} model, focusing on the \ac{jcs} aspect and introducing sparse \ac{md} reconstruction.
\modelname{} is presented in \secref{sec:approach}, along with a detailed explanation of each processing block. In \secref{sec:results}, \modelname{} is evaluated on a publicly available experimental dataset, showing its superior performance with respect to state-of-the-art solutions. Concluding remarks are provided in \secref{sec:conclusion}.

\section{Related Work}
\label{sec:rel-work}
\rev{
\subsection{micro-Doppler spectrogram applications}\label{sec:rel-mdapp}
\ac{md} analysis was originally introduced in the radar signal processing field~\cite{chen2000analysis, chen2006micro}, as an enabler for advanced target recognition and motion estimation applications~\cite{hanif2022micro, gerard2021micro, vandersmissen2018indoor}. In the past few years, the high sensitivity of \acp{mmwave} to \ac{md} shifts, together with \ac{dl} methods for spectrogram analysis and classification, has led to the successful utilization of \ac{md} analysis for human activity recognition~\cite{singh2019radhar, lai2021radar}, person identification~\cite{meng2020gait, pegoraro2021multiperson} and bio-mechanical gait analysis~\cite{seifert2019toward}. Due to its wide applicability to unobtrusive human motion analysis, \ac{md} processing has been identified as a prominent technology in remote healthcare and continuous monitoring of hospitals, homes, and public spaces.

Although we use a \ac{jcs} human activity recognition dataset to evaluate \modelname{}, our contribution is completely agnostic to the specific application scenario. As shown in \fig{fig:STAR_cover-diagram}, \modelname{} enables \ac{md} reconstruction in cases where 
very few, irregular channel estimates are available.
As such, it is an algorithm to \textit{obtain} the \ac{md}, rather than to analyze it, thus it applies to any type of target and motion after an appropriate training process.
}

\subsection{STFT-based micro-Doppler extraction.}\label{sec:rel-stft} \rev{The standard way of obtaining the \ac{md} signature of a target is to perform \ac{stft} on an estimate of the propagation channel~\cite{djurovic2017stft, vandersmissen2018indoor, gerard2021micro}.} 
In the context of \ac{jcs}, \cite{pegoraro2023rapid} has shown that similar processing can be performed using standard-compliant IEEE~802.11ay channels estimates, achieving comparable performance to radar devices in terms of activity recognition and person identification.

The drawback of these previous works is the need for \textit{regular} and \textit{dense} transmission of probing signals, to retrieve channel estimates with the required fine-grained Doppler resolution. This has a twofold negative impact on \ac{jcs} systems: (i)~it requires a considerable amount of power, time and frequency resources, and (ii)~it causes a significant overhead to the communication process, as dedicated sensing packets or waveforms have to be transmitted at the required rate, even though no data packets need to be transmitted by the users/applications. In the present work, instead, we focus on the less explored scenario where the above techniques fail because the channel estimates are obtained 
from incomplete sampling windows. \rev{Indeed, directly applying \ac{stft} on irregularly sampled data causes significant artifacts in the resulting spectrum. This poses the challenging research problem we tackle in this paper, involving the design of reconstruction algorithms that can recover the spectrum from a very few and irregularly sampled measurements. We do so by exploiting the channel sparsity in the frequency domain and the temporal correlation of \ac{md} spectra, stemming from the continuity of the underlying target movement.}

\subsection{Sparse micro-Doppler reconstruction.}\label{sec:rel-sparsemd}
In the radar context, sparse \ac{md} reconstruction with compressed sensing has been addressed in~\cite{li2017sparsity, li2014micro, stankovic2014compressive, sejdic2018compressive}.
In the \ac{jcs} literature instead, most works have focused on sparse sensing parameters estimation~\cite{kumari2021adaptive, zhang2017framework, li2017joint} but none dealt with the challenging problem of \ac{md} reconstruction, where all the Doppler frequencies of the different body parts have to be retrieved. To the best of our knowledge, only our previous work in~\cite{pegoraro2022sparcs} has directly tackled sparse \ac{md} reconstruction from IEEE~802.11ay channel estimates collected using irregular Wi-Fi traffic patterns. \rev{In the present work, we tackle this latter scenario due to its relevance to \ac{jcs}, where the irregularity or the channel measurements is a direct consequence of the underlying communication traffic patterns.} In addition, all the radar-based methods in~\cite{li2017sparsity, li2014micro, stankovic2014compressive, sejdic2018compressive} and~\cite{pegoraro2022sparcs} adopt standard iterative compressed sensing strategies, which suffer from two main limitations: (i)~the reconstruction accuracy is significantly degraded when the number of available measurements becomes very low, and (ii)~they are iterative methods requiring several iterations to converge. This leads to a consequent increase in processing delay and energy consumption, which hinders their real-time use in practical systems.
\rev{As a solution to the latter issues, \modelname{} enhances compressed sensing with data-driven feature learning, which exploits the sequential nature of \ac{md} to maintain a high reconstruction quality even when just a few measurements are available. This aspect is critical for \ac{jcs}, since communication traffic may present long idle periods and every additional transmission causes possible interference and overhead. Moreover, \modelname{} combines the concepts of attention~\cite{vaswani2017attention} and deep algorithm unrolling~\cite{monga2021algorithm} to reduce the number of iterations to just a single one. This drastic reduction in computational complexity makes our approach much more suited to real-time or near-real-time sensing applications.}

\subsection{Deep algorithm unrolling for sequential sparse recovery.} \label{sec:rel-dunroll} Deep \textit{unrolling} (or \textit{unfolding}) is a framework for enhancing traditional model-based iterative optimization algorithms with \ac{dl}. Each iteration of the algorithm is implemented by a neural network layer, whose parameters are learned from data through backpropagation~\cite{monga2021algorithm}. While most of the related research has focused on providing unrolled versions of popular iterative optimization algorithms, a few works have tackled the extension of deep unrolling to sparse temporal data~\cite{wisdom2017building, le2019designing, de2023designing}. These works dealt with the video recovery problems, modeling it as either an $\ell_1$-$\ell_1$ or an $\ell_1$-$\ell_2$ sparse reconstruction. The solver algorithm is unrolled using a \ac{rnn}, \cite{wisdom2017building, le2019designing}, or a Transformer network, \cite{de2023designing}. \rev{In previous unrolling formulations, the temporal correlation structure of the data was exploited to \textit{initialize} the sparse recovery solution at the algorithm onset, in \cite{wisdom2017building, le2019designing}, or at the beginning of every iteration, in \cite{de2023designing}, in an attempt of improving the recovery performance. In this work, we show that these techniques fail when the \ac{md} has to be reconstructed from very few measurements. In fact, initializing the \ac{iht} solution using past information simply provides an initial estimate of the support of the solution, which does not guarantee a good reconstruction. Conversely, \modelname{} makes full use of the signal temporal correlation to \textit{directly} suppress or enhance spectral components at the \textit{output} of the unrolled \ac{iht}, thus strengthening its reconstruction capability even when very few measurements are available. This grants it superior performance with respect to state-of-the-art approaches, as demonstrated by our experimental results in~\secref{sec:results}.}

\subsection{Deep learning for sinusoid estimation.} \label{sec:rel-sinus}
\ac{md} reconstruction is a sequential spectral estimation problem on time-varying sinusoidal signals. A few works have used deep learning methods to estimate the sinusoid parameters in noisy signals~\cite{izacard2019data, jiang2019deep}, and in low-resolution quantized signals~\cite{dreifuerst2022signalnet}.
These approaches rely on standard, \textit{blackbox} deep neural networks with little interpretability. \rev{\modelname{} instead combines an unrolled version of \ac{iht}, whose processing steps have a well-known interpretation of thresholded fixed point iterations~\cite{foucart2013mathematical}, and an attention layer that finds temporal correlations among \ac{md} spectra. This makes \modelname{} easily interpretable using signal processing domain knowledge.} Moreover, existing approaches significantly differ from our proposed technique in that: (i)~they only work for uniformly sampled signals, i.e., they can not handle incomplete sampling patterns and hence they are not suited for \ac{jcs}; (ii)~they are developed for single-shot sinusoidal spectral estimation, thus they can not exploit the sequentiality of \ac{md} data.
Conversely, \modelname{} is specifically designed to deal with incomplete sampling of the \ac{cir}, and it does so by jointly mimicking the structure of robust sparse reconstruction algorithms and exploiting the powerful feature extraction capabilities of attention mechanisms. This endows it with enhanced robustness to severe sparsity levels and a much-improved convergence speed.

\section{Background}
\label{sec:backg}
Next, we provide an overview of compressed sensing methods for sparse reconstruction and deep algorithm unrolling.

\subsection{Notation}\label{sec:notation}
\rev{A continuous-time signal $s$ is denoted by $s(\cdot)$, whereas square brackets are used for discrete-time signals, e.g., $s[\cdot]$.} $\Re (x)$ and $\Im (x)$ denote the real and imaginary part of $x \in \mathbb{C}$, while $|x|$ is its magnitude.
$\mathbf{X}^T$, $\mathbf{X}^H$, and $\mathbf{X}^*$ denote the transpose, Hermitian, and complex conjugate of matrix $\mathbf{X}$. $||\mathbf{x}||_p$ refers to the \mbox{$\ell_p$-norm} of vector $\mathbf{x}$, with $p=0, 1, 2$. 
The \textit{soft-thresholding} and \textit{hard-thresholding} operators applied to vector $\mathbf{x}$ are denoted by $\mathcal{S}_{\omega}(\mathbf{x})$ and $\mathcal{H}_{\omega}(\mathbf{x})$, respectively. Soft-thresholding is defined as $\mathcal{S}_{\omega}(\mathbf{x})=\text{sign}(\mathbf{x})\cdot \max(|\mathbf{x}|-\omega, 0)$, where operations on vector $\mathbf{x}$ are applied elementwise. $\mathcal{H}_{\omega}(\mathbf{x})$ sets to $0$ all the components of $\mathbf{x}$ but the $\omega$ largest ones.

When operating with \acp{nn}, complex-valued vectors $\mathbf{x} \in \mathbb{C}^{N}$ are transformed into real-valued vectors using the mapping $\mathbf{x}' = \mathcal{R}(\mathbf{x}) = \left[ \Re(\mathbf{x}) \; -\Im(\mathbf{x}) \right]^T \in \mathbb{R}^{2N}$.
When applied to matrices, the transformation computes \mbox{$\mathbf{X}' \in \mathbb{R}^{2M \times 2N}$} as
\begin{equation}
    \mathbf{X}' = \mathcal{R}(\mathbf{X}) = \left[\begin{array}{@{}cc@{}}
    \Re(\mathbf{X}) & -\Im(\mathbf{X})  \\
    \Im(\mathbf{X}) & \Re(\mathbf{X})  \\
  \end{array}\right].
\end{equation}
We denote by $\mathbf{F}_N$ the inverse $N$-point Fourier matrix, whose elements are $F_{n,m} = (1/\sqrt{N})\exp \left(j 2\pi nm/N\right), \,\, n,m = 0, \dots, N - 1$. $\mathbf{I}_N$ is the $N$-dimensional identity matrix. Finally, $\mathcal{U}([a, b])$ denotes the continuous uniform distribution in the interval $[a, b]$.

\subsection{Compressed sensing primer}
\label{sec:comp-sens}

\ac{cs} provides a framework to solve \textit{underdetermined} linear systems of the form
\begin{equation}\label{eq:lin-mod}
    \mathbf{x} = \mathbf{\Phi} \mathbf{z} + \mathbf{n},
\end{equation}
where $\mathbf{x}\in \mathbb{C}^M$ is a vector of noisy measurements, $\mathbf{z} \in \mathbb{C}^K$ is an unknown signal vector, to be reconstructed, $\mathbf{\Phi}\in \mathbb{C}^{M\times K}$ is the sensing matrix, and $\mathbf{n}\in \mathbb{C}^M$ is a noise vector.
Under the assumption that the signal vector is \textit{sparse}, meaning that $||\mathbf{z}||_0 \ll K$, i.e., it has only a few non-zero components, the core \ac{cs} result is that $\mathbf{z}$ can be reconstructed {\it exactly} even when $M < K$~\cite{eldar2012compressed, foucart2013mathematical}. 
This is subject to the requirement of having a sufficient number of measurements, that scales as the logarithm of $K$.   
The reconstruction is performed by solving an optimization problem of the form
\begin{equation}\label{eq:cs-prob}
\argmin_{\mathbf{z}} \left[ f(\mathbf{\Phi} \mathbf{z}, \mathbf{x}) + g(\mathbf{z})\right],
\end{equation}
where $f$ is a measure of the reconstruction error, e.g., the \mbox{$\ell_2$-norm} $||\mathbf{\Phi} \mathbf{z} - \mathbf{x}||_2$, and $g(\mathbf{z})$ is a regularization term that enforces the sparsity of the solution, e.g., $||\mathbf{z}||_0$ or $\lambda||\mathbf{z}||_1$, where $\lambda > 0$ is used to tune the importance of the regularization. Popular, fast iterative algorithms to solve \eq{eq:cs-prob} are: (i)~the \ac{ista}, that uses $g(\mathbf{z}) = \lambda ||\mathbf{z}||_1$, (ii)~\ac{iht}, that uses $g(\mathbf{z}) = ||\mathbf{z}||_0$, and (iii) \ac{omp}, which also uses the $\ell_0$ regularizer~\cite{eldar2012compressed}. \ac{ista} and \ac{iht} belong to the category of \textit{iterative thresholding} methods, whose $(i+1)$-th iteration computes
\begin{equation}\label{eq:thresh-sol}
    \mathbf{z}^{(i+1)} \leftarrow \mathcal{T}\left(\frac{1}{\mu}\mathbf{\Phi}^H\mathbf{x} + \left(\mathbf{I} - \frac{1}{\mu}\mathbf{\Phi}^H\mathbf{\Phi}\right)\mathbf{z}^{(i)}\right),
\end{equation}
where $\mu$ is the inverse of the learning step size. $\mathcal{T}$ is a suitable (algorithmic-dependent) thresholding operator. \ac{ista} uses soft-thresholding, $\mathcal{T}(\mathbf{z}) = \mathcal{S}_{\frac{\lambda}{\mu}}(\mathbf{z})$, while \ac{iht} uses hard-thresholding, $\mathcal{T}(\mathbf{z})=\mathcal{H}_{\Omega}(\mathbf{z})$, with $\Omega$ being a pre-defined sparsity level. Typical stopping criteria for \eq{eq:thresh-sol} involve setting a maximum number of iterations or stopping the algorithm upon convergence of the difference $||\mathbf{z}^{(i+1)} - \mathbf{z}^{(i)}||_2$.

\rev{
\subsection{Sequential compressed sensing}\label{sec:seq-cs}
\ac{cs} recovery has been extended to the case where the underlying signal is time-varying, and the processing relies upon a moving window approach~\cite{mota2016adaptive}. Denote by $\mathbf{x}[t]$ and $\mathbf{z}[t]$ the time-dependent vectors of measurements and unknown signal, respectively.
In the sequential formulation, the signal vectors in previous windows, $\mathbf{z}[t-1], \dots, \mathbf{z}[0]$, are used as \textit{side information} on the locations of the non-zero entries of $\mathbf{z}[t]$. \eq{eq:cs-prob} can be adapted to the sequential setting as follows~\cite{mota2016adaptive}
\begin{equation}\label{eq:cs-prob-seq}
    \argmin_{\mathbf{z}[t]} \left[ f(\mathbf{\Phi}_t\mathbf{z}[t], \mathbf{x}[t]) + d(\mathbf{z}[t], \dots, \mathbf{z}[0]) + g(\mathbf{z}[t]) \right],
\end{equation}
where $\mathbf{\Phi}_t$ is the sensing matrix in the $t$-th processing window, and function $d(\cdot)$ serves to regularize the optimization using past information.
Typically, $d(\cdot)$ is simplified to only capture the one-step correlation between $\mathbf{z}[t]$ and $\mathbf{z}[t-1]$, neglecting the potential dependency of $\mathbf{z}[t]$ on the full \ac{md} sequence. Previous works have then used $\ell_1$ or $\ell_2$ norm-based formulations for $d(\cdot)$ to aid the reconstruction of $\mathbf{z}[t]$, forcing it to share a similar support to $\mathbf{z}[t-1]$~\cite{mota2016adaptive, de2023designing}. Using the $\ell_1$-norm for $d(\cdot)$ and $g(\cdot)$, and the $\ell_2$-norm for $f(\cdot)$, the $\ell_1-\ell_1$ \ac{cs} problem is obtained~\cite{mota2016adaptive}, i.e., 
\begin{equation}\label{eq:l1-l1-prob}
    \argmin_{\mathbf{z}[t]} \left[ ||\mathbf{\Phi}_t\mathbf{z}[t]- \mathbf{x}[t]||^2_2 + \eta||\mathbf{z}[t] - \mathbf{D}\mathbf{z}[t-1]||_1 + \lambda||\mathbf{z}[t]||_1 \right],
\end{equation}
where $\eta>0$ weighs the importance of the past information, and matrix $\mathbf{D}$ represents a linear evolution model for the reconstructed signal across subsequent processing windows. The motivation behind modeling $d(\cdot)$ as in \eq{eq:l1-l1-prob} stems from a Laplacian approximation of the distribution of the error $\mathbf{z}[t] - \mathbf{D}\mathbf{z}[t-1]$, which holds well in video processing and other domains.
\eq{eq:l1-l1-prob} is typically solved using the algorithm in~\cite{mota2016adaptive} or an unrolled proximal gradient method~\cite{le2019designing} (see \secref{sec:unrolling} below).
}

\subsection{Deep unrolling for compressed sensing}
\label{sec:unrolling}

Although iterative algorithms have found widespread application thanks to their accuracy and ease of implementation, in practical settings they may require a large number of iterations to converge~\cite{monga2021algorithm}. The idea behind algorithm unrolling is to construct a \ac{nn} architecture in which each layer corresponds to one iteration of \eq{eq:thresh-sol}. This is done by exploiting the structural similarity between \eq{eq:thresh-sol} and a recurrent \ac{nn} layer with state vector $\mathbf{z}^{(i)}$ and a fixed input $\mathbf{x}$, where the thresholding operator plays the role of a non-linear activation function~\cite{goodfellow2016deep}. In the unrolling implementation, \eq{eq:thresh-sol} is usually rewritten by using the complex-to-real transformation, $\mathcal{R}(\cdot)$, introduced in \secref{sec:notation}. 
The input-output equation of an unrolled \ac{nn} layer can be written as
\begin{equation}\label{eq:unroll-layer}
    \mathbf{z}^{(i+1)} = \mathcal{T}\left(\frac{1}{\mu}\mathbf{W}^T\mathbf{x} + \mathbf{S}\mathbf{z}^{(i)}\right),
\end{equation}
where $\mathbf{W}$ is a set of learnable weights and $\mathbf{S}$ can be selected as in the original algorithm, i.e., $\mathbf{S} = \mathbf{I} - \frac{1}{\mu}\mathbf{W}^T\mathbf{W}$, or as an additional set of learnable weights~\cite{monga2021algorithm}.
The network is constructed by stacking a fixed number of layers of the type in \eq{eq:unroll-layer}, with weights $\mathbf{W}, \mathbf{S}$ being \textit{shared} among the different layers. Training is then performed with standard backpropagation on a dataset of $D$ input-output pairs, $\mathcal{X} = \{\mathbf{x}_m, \mathbf{z}_m^*\}^{D}_{m=1}$, where $\mathbf{z}_m^*$ is the solution obtained with the original iterative algorithm with input $\mathbf{x}_m$, which can be pre-computed. 
In the literature, unrolled versions of several popular algorithms have been proposed~\cite{lohit2019unrolled, hosseini2020dense}. The unrolled \ac{ista} and \ac{iht} algorithms are commonly referred to as \ac{lista} and \ac{liht}~\cite{gregor2010learning, xin2016maximal}.

\rev{Reference~\cite{le2019designing} has shown that deep unrolling is well suited to address sequential \ac{cs} problems tackled via, e.g., $\ell_1-\ell_1$ minimization. There, an \ac{rnn} is presented that unrolls a proximal gradient method solving \eq{eq:l1-l1-prob}. The key difference between this approach and \modelname{} lies in the way in which the prior information, represented by past solutions $\mathbf{z}[t-1], \dots, \mathbf{z}[0]$, is exploited. In detail, instead of specifying a model for $d(\cdot)$, we let the \ac{nn} learn it directly from data through an attention layer. This better exploits the previous reconstructions, which ultimately leads to enhanced results with very few available measurements.}

\section{Channel model and micro-Doppler}
\label{sec:sparse-md}

In \ac{jcs} systems, \ac{cir} estimates are reused (besides using them to decode data) to obtain information about the sensing targets. 
In the interest of better conveying our framework, we keep the experimental setup simple, by assuming to be in a \textit{monostatic} scenario, i.e., with co-located transmitter and receiver. We underline that this is not a restrictive assumption for \modelname{}. In this setup, the channel can be estimated whenever the reflections of transmitted packets are collected back at the transmitter. Therefore, \ac{cir} estimates are obtained at irregular (and random) time instants which depend on the transmission pattern at the transmitter side, and coincide with the reception of packets. As shown in~\cite{pegoraro2022sparcs}, \ac{cir} can sampled at \textit{regular} intervals, obtaining a grid of channel samples evenly spaced by $T_c$ seconds, where $T_c$ is the (arbitrary) channel sampling period. However, in communication networks, it is very unlikely that data packets are available for transmission every $T_c$ seconds (leading to channel estimates at this granularity), and this very much depends on the application data pattern. Thus, it follows that the resulting \ac{cir} measurements grid is likely to be incomplete, i.e., some of the measurements on the grid are missing. To compensate for this, one may think of transmitting dummy packets (with no data) to attain channel estimates at regular intervals, but this is undesirable as it implies a large communication overhead.

\subsection{Channel impulse response model}\label{sec:cir-model}
 
\rev{In the following, we use a \ac{sc} \ac{cir} model similar to that in~\cite{pegoraro2022sparcs}, since the dataset used for our experimental validation is based on a \ac{sc} IEEE~802.11ay \ac{jcs} implementation \cite{pegoraro2022disc}. However, the presented approach is equally applicable to \ac{ofdm} waveforms.

For the channel model, we consider discrete timesteps $nT_c, n \in \mathbb{Z}^+$, with time granularity $T_c$. Although our system operates on incomplete windows of channel samples, the \ac{cir} is mathematically modeled as if all the samples were available. Shortly below, in \secref{sec:md-recovery}, we show how the missing samples can be added to the formulation. At time $nT_c$, the \ac{cir} is a function of the propagation delay $\tau$, expressed as a sum of $L$ Dirac delta components which correspond to the \textit{resolvable} signal propagation paths~\cite{zhang2021overview}. 
The finite delay resolution of the system is \mbox{$\Delta \tau = 1/B$}, where $B$ is the transmitted signal bandwidth. Denoting by $\tau_l$ the propagation delay of the \mbox{$l$-th} scatterer, and by $h_l(nT_c)$ the \mbox{$l$-th} complex \ac{cir} component, we have 
\begin{equation}\label{eq:cir-continuous}
    h(\tau, n T_c) = \sum_{l=1}^{L} h_{l}(nT_c) \delta(\tau - \tau_l (nT_c)).
\end{equation}
The propagation delay of path $l$ is associated with a specific distance from the \ac{jcs} transceiver, according to the relation $d_{\ell}=c\tau_l/2$, where $c$ is the speed of light. Hence, the signal bandwidth determines a minimum distance threshold, equal to $\Delta d = c\Delta\tau/2$, under which two targets produce reflections that overlap in the same \ac{cir} peak at the receiver. For complex targets with many moving parts, such as the human body, the system bandwidth in common communication systems (even in the \ac{mmwave} range) is typically insufficient to entirely resolve the resulting reflections~\cite{pegoraro2022sparcs}. 

Denote by: $Q_l(nT_c)$ the number of reflections due to the unresolvable scatterers composing the sensing target, $f_{l, q}$ the Doppler shift of the $q$-th of such reflections and $\alpha_{l,q}$ a complex coefficient that accounts for the scatterer's \ac{rcs}, the propagation loss, and the beamforming gains.
The \mbox{$l$-th} \ac{cir} component at time $nT_c$ is
\begin{equation}\label{eq:cir}
    h_l(nT_c) = \sum_{q=1}^{Q_l(nT_c)} \alpha_{l,q}(nT_c)  e^{j2\pi f_{l, q}(nT_c) nT_c}.
\end{equation}
The expression of the Doppler shift depends on the movement speed of the scatterer, denoted by $v_{l, q}$, on the angle of the direction of motion with respect to the incident wave, $ \theta_{l, q} $, and on the carrier frequency $f_c$, as 
\begin{equation}\label{eq:doppler-shift}
    f_{l, q}(nT_c) = \frac{v_{l, q}(nT_c) \cos \theta_{l, q}(nT_c) }{c} f_c ,
\end{equation}
where $c$ is the speed of light.

Finally, we make the two following assumptions to further simplify \eq{eq:cir}.

\textit{Assumption 1 - Window-based processing}: The \ac{cir} samples are processed in time windows spanning $K$ subsequent timesteps, where $KT_c$ is sufficiently short, so that the parameters of the scatterers, $\alpha_{l, q}, f_{l,q}, Q_l$, can be considered constant within each window. Timesteps within each processing window are indexed by variable $k=1,\dots, K$. We allow instead the parameters to change across different processing windows (indexed by $t$). This assumption is ubiquitous in radar signal processing and \ac{jcs}, and has been empirically verified for human sensing~\cite{zhang2021overview, kumari2017ieee, vandersmissen2018indoor}.}

\textit{Assumption 2 - Tracking multipath reflections}: We assume we can track the signal propagation paths corresponding to the sensing targets of interest, following the evolution of their delay parameter $\tau_l$ across time. \rev{Tracking $\tau_l$ amounts to tracking the distance of the target from the transmitter, termed $d_l$, across time, since it holds that $d_l = c\tau_l /2$. This is a standard processing step in wireless human sensing applications, which can be done using one of the many existing multiple target tracking techniques based on multiple Kalman filters~\cite{zhang2021overview}, probabilistic data association filters~\cite{shalom2009probabilistic}, or multiple hypotheses tracking~\cite{reid1979algorithm}. The performance of these algorithms depends on the system bandwidth, $B$, which determines the delay resolution, and on the extension of the target. Tracking of several human targets in challenging indoor scenarios has been successfully demonstrated in many works, e.g.,~\cite{pegoraro2023rapid, zhao2019mid}.
Tracking $\tau_l$ allows extracting the \ac{cir} component due to the sensing target, separating it from those of the other scatterers. As such, it is a necessary processing step for any wireless sensing system that obtains target-specific \ac{md} signatures: If no separation of the different target reflections were performed, the \ac{md} contributions of different targets would overlap in the Doppler domain, interfering with one another. It follows that Assumption 2 is not due to a limitation of the proposed system but to an intrinsic requirement for multitarget wireless sensing. Note that in single-target scenarios Assumption 2 becomes unnecessary and it can be removed.}

\rev{Combining assumptions $1$ and $2$ allows rewriting \eq{eq:cir} by replacing the time dependency of $\alpha_{l,q}, f_{l,q}, Q_l$ with a coarser dependency on the processing window, $t$, and dropping index $l$ as we consider each target to be consistently tracked and separated from the others. Hence, denoting by $\alpha_{q}[t], f_{q}[t], Q[t]$ the sensing parameters of target $q$ in the $t$-th window, \eq{eq:cir} becomes 
\begin{equation}\label{eq:cir-simple}
    h[t, k] =\sum_{q=1}^{Q[t]} \alpha_q[t] e^{j2\pi f_{q}[t] (tK + kT_c)},
\end{equation}
where we use the compact, discrete-time notation $h[t, k]$ to refer to the $k$-th \ac{cir} sample in the $t$-th processing window.
\eq{eq:cir-simple} forms the basis for the formulation of the sparse sequential \ac{md} recovery problem, as it expresses each \ac{cir} component as a superposition of $Q[t]$ complex sinusoids with frequencies $f_q[t]$. Each sinusoid is associated with one of the scatterers overlapping in the same \ac{cir} component. 
The \ac{cir} can be considered sparse in the frequency domain if $Q[t] \ll K$. This follows from the fact that each complex exponential has a single active frequency, hence the Fourier transform of \eq{eq:cir-simple} has few non-zero elements. In \cite{pegoraro2022sparcs}, it has been shown that for targets involving few moving parts (such as the human body), channels observed by wideband, directional communication systems lead to few scattering components $Q[t]$, hence justifying the use of \ac{cs} methods.

We stress that \eq{eq:cir-simple} refers to a single component in the \ac{cir}, but contains information about multiple unresolvable paths that all overlap in the same \ac{cir} peak. This demonstrates the importance of \ac{md} analysis in recognizing the movement of complex targets, as it allows resolving multiple moving parts in the Doppler domain, leveraging their different moving speeds. Moreover, the evolution of the sensing parameters, $\alpha_{q}[t], f_{q}[t], Q[t]$, across different windows is \textit{correlated}. This stems from the complex dynamics of the underlying movement of the target and how these affect the channel depending on the observation angle, distance, and propagation characteristics. As a result, such sequential correlation is extremely challenging to model, as further discussed in \secref{sec:seq-md-rec}.
}

\subsection{\rev{Sparse sequential micro-Doppler reconstruction}}\label{sec:md-recovery}

\rev{The \ac{md} signature of a target is typically obtained by using the \ac{stft} \cite{chen2006micro, vandersmissen2018indoor, pegoraro2023rapid},  which applies a \ac{dft} to each window of $h[t, k]$. The resulting spectrum peaks at values $f_q[t] = v_q[t] \cos \theta_q[t] f_c/  c$. Hence, spectral analysis of the \ac{cir} reveals important features of the underlying physical movement of the target, especially when considering its temporal evolution across subsequent windows. }
The frequency resolution ($\Delta f$) and the maximum resolvable frequency ($f_{\rm m}$) of the \ac{stft} depend on $T_c$ as $\Delta f = 1/(KT_c)$, and $f_{\rm m} = 1/(2T_c)$, respectively. These can then be mapped onto the corresponding velocity of the scatterer as $\Delta v = c/(2f_c  K T_c)$ and $v_{\rm m} = c/(4f_c T_c)$.
When \ac{cir} measurements are incomplete, \ac{stft} is not directly applicable as the resulting spectrum would be degraded by the lack of a fixed sampling interval. Hence, alternative approaches have to be sought relying on {\it sparse reconstruction}~\cite{pegoraro2022sparcs}. 

\subsubsection{Sparse \ac{md} reconstruction} Due to the missing \ac{cir} values, out of $K$ timesteps in window $t$ we only have $M_t$ available samples, whose indices in the window are denoted by $m_1, \dots, m_{M_t}$.
\rev{We define vector $\mathbf{h}[t] = [h[t, m_1], \dots, h[t, m_{M_t}]]^T$, containing the available \ac{cir} samples in the window, and 
matrix $\mathbf{M}_t=\left[ \mathbf{e}_{m_1}, \dots, \mathbf{e}_{m_{M_t}}\right]^T$, 
where $\mathbf{e}_{i}$ is the $i$-th vector of the canonical basis.} Left multiplication of a matrix by $\mathbf{M}_t$ has the effect of selecting the rows of such matrix whose indices correspond to the available samples. Moreover, denote by $\mathbf{z}[t] \in \mathbb{C}^{K}$ the \ac{dft} of the (unknown) complete window of \ac{cir} samples $h[tK], \dots, h[(t+1)K - 1]$.
The following compressed sensing model can be formulated relating $\mathbf{h}[t]$ and $\mathbf{z}[t]$,
\begin{equation}\label{eq:cs-model}
    \mathbf{h}[t] =  \mathbf{M}_t  \mathbf{F}_K\mathbf{z}[t] + \mathbf{n},
\end{equation}
where $\mathbf{n}$ is a $M_t$-dimensional complex noise vector, and $\mathbf{F}_K$ is the inverse Fourier matrix defined in \secref{sec:notation}.
By setting $\mathbf{\Phi}_t = \mathbf{M}_t  \mathbf{F}_K$, \eq{eq:cs-model} is in the form expressed by \eq{eq:lin-mod}, hence we can tackle it by solving the optimization problem in \eq{eq:cs-prob}.
Our aim is to recover $\mathbf{z}[t]$ from the incomplete measurement vector $\mathbf{h}[t]$, so that we can obtain the \ac{md} spectrum of the $t$-th \ac{cir} window as $\mathbf{y}[t] =|\mathbf{z}[t]|^2$. 
To this end, we solve the following compressed sensing problem, which finds a vector $\mathbf{z}[t]$ which is a solution to \eq{eq:cs-model}, while being at least $\Omega$-sparse
\begin{equation}\label{eq:cs-iht}
    \argmin_{\mathbf{z}[t]} ||\mathbf{h}[t] - \mathbf{\Phi}_t\mathbf{z}[t]||_2^2 \, \mbox{ s.t. } ||\mathbf{z}[t]||_0  \leq \Omega.
\end{equation}
The constant $\Omega \in \mathbb{N}$ is a pre-defined sparsity level that is closely related to the value of $Q$ in \eq{eq:cir-simple}. We set $\Omega$ as an upper bound to $Q$, as the latter is unknown in practice. In this way, the sparse reconstruction retrieves a solution that is $\Omega$-sparse in the frequency domain, ensuring that all the $Q$ frequency components of the \ac{cir} can be reconstructed. To solve \eq{eq:cs-iht}, previous work has adopted \ac{iht}~\cite{pegoraro2022sparcs}. However, this solution has significant limitations in terms of computational speed, as it involves an iterative process that may take several iterations to converge, and suffers from low reconstruction quality when very few measurements per window are available~\cite{pegoraro2022sparcs}.

\rev{\subsubsection{Sequential \ac{md} reconstruction}\label{sec:seq-md-rec}
Solving \eq{eq:cs-iht} independently for each processing window does not fully exploit the temporal features of the \ac{md} spectrogram. Indeed, the evolution of the \ac{md} in time exhibits particular patterns that depend on the type of target and movement. In our model, this can be expressed as the availability of \textit{side information} on $\mathbf{z}[t]$, which depends on the previous \ac{md} spectrum windows. This can be modeled as a sequential \ac{cs} problem, as discussed in \secref{sec:seq-cs}. \eq{eq:cs-iht} is modified to account for the correlation among \ac{md} windows as follows \cite{mota2016adaptive}
\begin{equation}\label{eq:cs-seq-iht}
    \begin{aligned}
        \argmin_{\mathbf{z}[t]} & \left[ ||\mathbf{h}[t] - \mathbf{\Phi}_t\mathbf{z}[t]||_2^2\, +\, d(\mathbf{z}[t], \dots, \mathbf{z}[0]) \right] \\ 
    \mbox{ s.t. } & ||\mathbf{z}[t]||_0  \leq \Omega.
    \end{aligned}
\end{equation}
Using the 
formulation in \eq{eq:l1-l1-prob} to solve the above problem fails to capture complex and long-term correlations among the sequence of \ac{md} spectra. Moreover, as we discuss in \secref{sec:SOTA_comparison}, previous approaches that adopted unrolling to address sequential \ac{cs} did not fully exploit the information from previous reconstructions, as such information is used to obtain a better \textit{initialization} for the reconstruction algorithm (at the input of the \ac{iht} processing block). 

In the next section, we describe our approach, which is significantly less computationally complex than \ac{iht} and fully exploits the \ac{md} temporal correlation structure. To reduce the computational complexity of standard \ac{iht}, we unroll a \textit{single iteration} of \ac{iht} into a neural network. Then, we enhance the resulting solution by learning sequential features of the \ac{md} using an attention mechanism, which acts as a refinement step applied to the output of the \ac{iht} block. This avoids having to specify a model for the function $d(\cdot)$ in \eq{eq:cs-seq-iht} and rather lets the neural network learn \ac{md} correlations directly from data.}

\section{The \modelname{} architecture}\label{sec:approach}

\begin{figure}
    \centering
    \includegraphics[width= \linewidth]{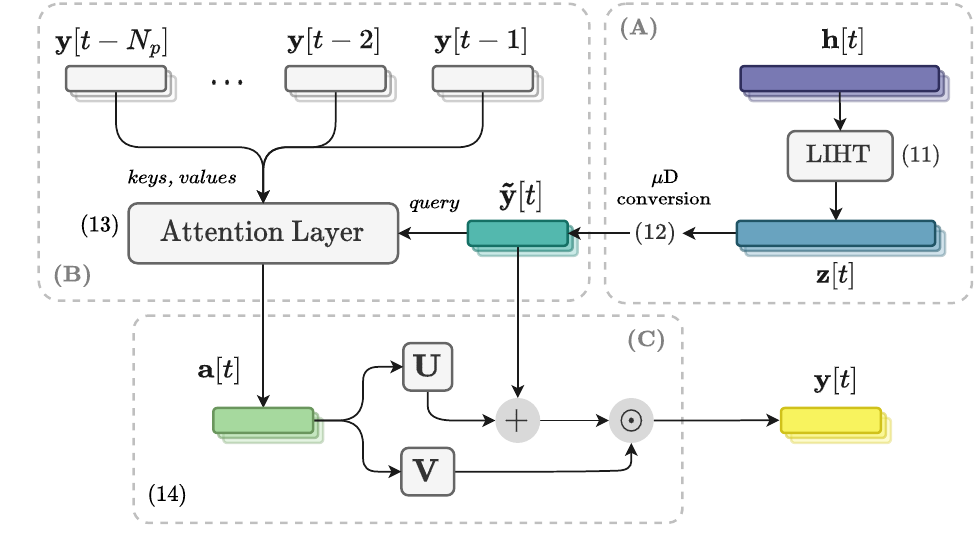}
    \caption{Block diagram of \modelname{}.}
    \label{fig:model-diagram}
\end{figure}

The block diagram of \modelname{} is shown in \fig{fig:model-diagram}, while the step-by-step computations are reported in \alg{alg:our-liht}. \modelname{} processes the current input measurements by unrolling the \ac{iht} algorithm, and subsequently improves the resulting reconstruction by leveraging the sequential structure of the \ac{md} spectrogram.  
The computations performed by \modelname{} can be subdivided into three distinct blocks:

\noindent \textbf{(A) Single-layer \ac{liht}:} With the first block, we retrieve an approximate solution for the reconstruction of the micro-Doppler window at time $t$ (of size $K$), denoted by $\tilde{\mathbf{y}}[t]$, with minimal computations. To do so, the incomplete \ac{cir} input window is processed by a single \ac{liht} layer, which unrolls one \ac{iht} iteration, as discussed in-depth in \secref{sec:LIHT-module}. The \ac{liht} output is then refined in steps $(2)$ and $(3)$. 

\noindent \textbf{(B) Attention mechanism:} In this block, temporal features of the \ac{md} are exploited to provide a \textit{context} for the approximate solution found in $(1)$. \ac{md} signatures of human movements exhibit a strong temporal correlation, which we take into account to refine the quality of the reconstruction. To this end, we introduce a scaled dot product attention mechanism. This computes the correlation of the current approximate reconstruction with past recovered \ac{md} spectra, to obtain a context feature vector $\mathbf{a}[t]$. Details regarding the attention mechanism are reported in \secref{sec:att-module}.

\noindent \textbf{(C)  Solution refinement:} Finally, the outputs from the previous two blocks are combined to retrieve the final sparse \ac{md} spectrum. The solution refinement block learns a transformation of the current approximate solution, $\tilde{\mathbf{y}}[t]$, based on the context feature vector $\mathbf{a}[t]$. This is achieved by stacking two parallel feedforward \ac{nn} layers (with weights $\mathbf{U}$ and $\mathbf{V}$) that are fed with $\mathbf{a}[t]$, and whose output is used to transform $\tilde{\mathbf{y}}[t]$ additively and multiplicatively. The resulting output is the current \ac{md} reconstruction. Block \textbf{(C)} is thoroughly discussed in \secref{sec:sol-improvement-module}.

\subsection{LIHT Module}
\label{sec:LIHT-module}
The \ac{liht} block operates on each incomplete \ac{cir} measurement window independently, as summarized in lines~$3-5$ of \alg{alg:our-liht}. 
We use a single \ac{liht} layer to obtain a baseline sparse reconstruction of $\mathbf{z}[t]$. Calling $\mathcal{H}_{\Omega}$ the hard-thresholding operator described in \secref{sec:comp-sens}, $\mu$ the reciprocal of the learning step size, and initializing the reconstruction as $\mathbf{z}^{(0)}[t] = \mathcal{H}_{\Omega} \left( \frac{1}{\mu} \mathbf{W}^T \mathbf{h}[t] \right)$, the \ac{liht} block computes
\begin{equation}\label{eq:liht-module}
    \mathbf{z}[t] = \mathcal{H}_{\Omega} \left ( \left( \mathbf{I} - \frac{1}{\mu} \mathbf{W}^T\mathbf{W}\right) \mathbf{z}^{(0)}[t] + \frac{1}{\mu} \mathbf{W}^T \mathbf{h}[t] \right ).
\end{equation}
The specific choice of the initialization follows~\cite{gregor2010learning}.
All vectors and matrices in \eq{eq:liht-module} are subjected to the complex-to-real transformation defined in \secref{sec:notation}. Therefore, we have $\mathbf{h}[t], \mathbf{z}[t] \in \mathbb{R}^{2K}$.
Matrix $\mathbf{W}$ is initialized as $\mathbf{W} = \mathcal{R}(\mathbf{M}_t\mathbf{F}_K)\in \mathbb{R}^{2M_t \times 2K}$ and then learnt during training. 

We remark that $\mathbf{z}[t]$ is a reconstruction of the \ac{dft} of $\mathbf{h}[t]$, not of its \ac{md} spectrum $\mathbf{y}[t] \in \mathbb{R}^{K}$. To obtain the latter, we compute
\begin{equation}\label{eq:md-conversion}
    \tilde{\mathbf{y}}[t] = \left[\begin{array}{@{}cc@{}}
    \mathbf{I}_K & \mathbf{I}_K  \\
  \end{array}\right] \mathbf{z}[t]^2,
\end{equation}
where the square operation is applied elementwise to $\mathbf{z}[t] \in \mathbb{R}^{2K}$. In detail, with \eq{eq:md-conversion} we obtain the squared magnitude (\ac{md} spectrum) of the DFT of the $t$-th \ac{cir} window. The two identity matrices are required to combine the real and imaginary part of such (complex-valued) DFT, that respectively appear in the first and second $K$ elements of vector $\mathbf{z}[t]$ (due to our definition of the transform $\mathcal{R}$).

\subsection{Attention mechanism}
\label{sec:att-module}

This is a key (and novel) step of the proposed technique: As it will be discussed in \secref{sec:SOTA_comparison}, positioning the attention layer as we did, i.e., towards the end of the processing chain, is non-standard and we found it to be highly effective. To exploit the strong temporal correlations in the \ac{md} spectrograms due to the sequentiality of human movement, \modelname{} looks for correlations between $\tilde{\mathbf{y}}[t]$, and the reconstructions obtained at previous time steps, $\mathbf{y}[t-1], \mathbf{y}[t-2], \dots, \mathbf{y}[t-N_p]$. $N_p$ is the number of past \ac{md} windows considered in the computation of the context features, which is set as a hyperparameter of our model. The key idea behind this attention layer is to obtain context features to be used by block \textbf{(C)} to improve the current approximate solution $\tilde{\mathbf{y}}[t]$. Specifically, the attention layer learns the temporal correlation model that best represents the evolution of a spectrogram window.\footnote{This model is strongly connected with the underlying physical movement that is being tracked, e.g., human-related.} Such model is then utilized to refine $\tilde{\mathbf{y}}[t]$, improving the reconstruction of those window elements that are heavily corrupted due to undersampling. 
\rev{Note that our approach does not explicitly require specifying function $d(\cdot)$ in \eq{eq:cs-seq-iht}, but rather uses an attention mechanism to learn complex sequential features of the \ac{md}.} The computations performed by the attention mechanism are described next, and reported in lines~$6-7$ of \alg{alg:our-liht}. 

Following the original terminology introduced in~\cite{vaswani2017attention}, attention compares a set of \textit{query} vectors to some \textit{key} vectors, according to a specific alignment function, to obtain a set of \textit{attention weights} for each query vector. Such weights are then mapped within $[0, 1]$ by a $\mathrm{Softmax}$ function and used to perform a weighted convex sum of some \textit{value} vectors, which represents the final output of the attention layer. 
In \modelname{}, we define matrix \mbox{$\mathbf{Y}[t] = \left[ \mathbf{y}[t-1], \dots, \mathbf{y}[t-N_p] \right]^T$}, which contains the reconstructions obtained at past time steps. The columns of $\mathbf{Y}[t]$ serve as both the keys and the values of our attention mechanism. Queries are instead represented by the current approximate reconstruction $\tilde{\mathbf{y}}$, which we want to contextualize with respect to the sequence of \ac{md} spectra. As the alignment function, we use the dot product, which measures the level of correlation between queries and the key while maintaining the whole layer extremely efficient and lightweight. The computations performed by the attention mechanism are summarized in the following equation
\begin{align}\label{eq:attention}
    \nonumber \mathbf{a}[t] &= \sum_{i = 1}^{N_p} \frac{e^{\frac{1}{\sqrt{K}} \mathbf{y}[t-i]^T \tilde{\mathbf{y}}[t]}}{\sum_{u=1}^{N_p} e^{\frac{1}{\sqrt{K}} \mathbf{y}[t-u]^T \tilde{\mathbf{y}}[t]}}  \mathbf{y}[t-i]\\
    &= \mathbf{Y}[t]^T \mathrm{Softmax}\left( \frac{1}{\sqrt{K}} \mathbf{Y}[t] \tilde{\mathbf{y}}[t] \right).
\end{align}
In \eq{eq:attention}, the correlations are computed directly in the frequency spectrum domain, without introducing any additional learnable parameter. This is in contrast with the typical approach used in \ac{dl} models of projecting queries, keys, and values to high dimensional feature spaces before the computation of the attention weights~\cite{vaswani2017attention}. For the considered human-sensing application, experimental results showed no performance improvement when learning such projections.

\subsection{Solution refinement Module}
\label{sec:sol-improvement-module}
In this final step, the initial reconstruction $\mathbf{z}[t]$ is refined using the context provided by the attention layer. 
The key idea behind this step is to leverage the typical structure of \ac{md} spectra, which show \textit{localized} spectral components around the Doppler frequencies of the different body parts. 

When applying the standard \ac{iht} with few input measurements we observe two main shortcomings (see \fig{fig:md-comparison} and \fig{fig:md-comparison-hands}): (i)~the resulting reconstruction shows unstructured background noise with no physical meaning, and (ii)~the frequency components due to the body parts are inaccurately reconstructed.
Therefore, we apply two different kinds of operations to $\tilde{\mathbf{y}}[t]$, in order to mitigate these two issues separately:
\begin{enumerate}
    \item First, an \textit{additive} transformation is applied to $\tilde{\mathbf{y}}[t]$. This is done to give the model the ability to learn how to improve the solution in those frequency bins that have been poorly reconstructed. The additive term is parametrized as a dense neural network layer with a \ac{relu} activation function, to enforce only positive-valued changes to the approximate known solution. The \ac{relu} is defined as $\mathrm{ReLU}(x) = \max(0, x)$;
    \item Next, \modelname{} applies a \textit{multiplicative} masking to the output of step $1)$. This step is designed to let the model learn how to denoise the reconstruction. The multiplicative term is parametrized by a dense neural network layer with a sigmoid activation function, denoted by $\sigma(\cdot)$, to constrain the output in the interval $[0, 1]$.
\end{enumerate}
The computations performed by the refinement block are (line~$8$ in \alg{alg:our-liht})
\begin{equation}\label{eq:sol-refinement}
    \mathbf{y}[t] = \left( \mathbf{z}[t] + \mathrm{ReLU}\left( \mathbf{U} \mathbf{a}[t] + \mathbf{b} \right)\right) \odot \sigma\left( \mathbf{Va}[t] \right),
\end{equation}
with $\mathbf{y}[t]$ being the final output \ac{md} spectrum.

\begin{algorithm}[t!]
\setstretch{1.2}
\small
	\caption{Computational steps of \modelname{}.}
	\label{alg:our-liht}
	\begin{algorithmic}[1]
		\REQUIRE Sequence of incomplete CIR windows \mbox{$\mathbf{h}[t], t=1, 2, \dots$}
		\ENSURE \ac{md} spectrogram \mbox{$\mathbf{y}[t], t=1, 2,\dots$}.

\STATEx \textcolor{gray}{\texttt{// Initialize learnable weights}}

\STATE $\mathbf{W} \leftarrow \mathbf{F}_K$;
    $\mathbf{U, V, b} \sim \mathcal{U}(\frac{1}{\sqrt{K}},\frac{1}{\sqrt{K}})$
   
            \FOR{$t=1, 2, \dots$} 
            \STATEx \textcolor{gray}{\texttt{// Single-layer \ac{liht}}}
            \STATE $\mathbf{z}^{(0)} \leftarrow \mathcal{H}_{\Omega} \left( \frac{1}{\mu} \mathbf{W}^T \mathbf{h}[t] \right) $
            \STATE $\mathbf{z}[t] \leftarrow \mathcal{H}_{\Omega} \left\{ \left( \mathbf{I} - \frac{1}{\mu} \mathbf{W}^T\mathbf{W}\right) \mathbf{z}^{(0)}[t] + \frac{1}{\mu} \mathbf{W}^T \mathbf{h}[t] \right\}$
            \STATE $\tilde{\mathbf{y}}[t] \leftarrow \left[\begin{array}{@{}cc@{}}
                \mathbf{I}_K & \mathbf{I}_K  \\
            \end{array}\right] \mathbf{z}[t]^2$ 
            \STATEx \textcolor{gray}{\texttt{// Attention mechanism}}
            \STATE $\mathbf{Y}[t]_i \leftarrow \left[ \mathbf{y}[t-i]^T \right]$

		\STATE $\mathbf{a}[t] \leftarrow \mathbf{Y}[t]^T \mathrm{Softmax}\left( \frac{1}{\sqrt{K}} \mathbf{Y}[t] \tilde{\mathbf{y}}[t] \right)$
        \STATEx  \textcolor{gray}{\texttt{// Solution refinement}}
		\STATE $\mathbf{y}[t] \leftarrow \left( \tilde{\mathbf{y}}[t] + \mathrm{ReLU}\left( \mathbf{U} \mathbf{a}[t] + \mathbf{b} \right)\right) \odot \sigma\left( \mathbf{Va}[t] \right)$
		\ENDFOR
		
	\end{algorithmic}
\end{algorithm}

\rev{
\subsection{Final remarks}\label{sec:remarks}
In this section, we provide some final remarks to link \modelname{} to the mathematical formulation of the problem defined in~\secref{sec:seq-md-rec}, and to other sequential \ac{cs} algorithms. 

\modelname{} solves the sequential \ac{cs} problem in \eq{eq:cs-seq-iht} taking a different approach compared to existing works, such as \ac{slista}~\cite{wisdom2017building}, the $\ell_1 - \ell_1$ \ac{rnn} in~\cite{le2019designing}, and \ac{dust}~\cite{de2023designing}. These approaches propose \acp{nn} that maintain the same structure of the original iterative optimization algorithm and exploit the temporal correlation of the data to initialize the solver. Specifically, the unrolling formulation in~\cite{wisdom2017building,le2019designing} uses past information to initialize the solution once, before the first iteration of the unrolled algorithm. \cite{de2023designing} instead uses it to initialize the solution at the beginning of every iteration. In \secref{sec:results}, we show that with highly incomplete measurements (i.e., $90$\% missing samples) these initialization strategies fail, as the information carried by the current measurement vector can be very low, making the unrolled iterative algorithm converge to a poor solution despite the good initialization. 

To cope with this, with \modelname{} we take an opposite approach with respect to existing works. We use a single deep unrolling iteration to find an initial solution, which is then {\it refined} by exploiting the past evolution of the \ac{md} using temporal features obtained from an attention layer. This avoids a direct optimization of \eq{eq:cs-seq-iht}, which requires specifying function $d(\cdot)$ and, in turn, making restrictive assumptions on the \ac{md} dynamics. Thus, the attention layer is used to \textit{learn} how to enhance the output of the sparse recovery problem solved through the \ac{liht} block.
It follows that \modelname{} splits the problem into a \ac{cs} part, implemented via deep unrolling, and a sequential feature learning part, implemented through the attention mechanism. The features obtained from the attention layer are then utilized to refine the solution {\it at the output} of the unrolling block, rather than to initialize it. 
Note that, although the attention layer is not part of the model-based unrolling framework (block \textbf{(A)}), its effect on the current solution $\tilde{\mathbf{y}}[t]$, by exploiting the correlation structure learned from past outputs, is still clearly interpretable.
}

\section{Experimental Results}\label{sec:results}

In this section, we present experimental results to validate \modelname{} and compare it against state-of-the-art algorithms. The code implementation of our model, which was carried out in PyTorch~\cite{paszke2019pytorch}, will be made available on GitHub\footnote{\texttt{\url{https://github.com/rmazzier/STAR}}} to facilitate reproducibility. 

\subsection{Dataset description and system parameters}
\label{sec:dataset}
\begin{table}[t!] 
\footnotesize
	\caption{\ac{cir} parameters of the DISC dataset.} \label{tab:disc-params}
    \vspace{-0.25cm}
	\begin{center}
		\begin{tabular}{cccccc}
			\toprule	
			$B$~[GHz]&$\Delta \tau$~[ns]&$T_c$ [ms]&$f_c$~[GHz]&$\Delta v$~[m/s]&$v_{\rm m}$ [m/s]\\
            \midrule
			$1.76$&$0.568$&$0.27$&$60$&$0.14$&$\pm 4.48$\\
			\bottomrule
		\end{tabular} 		
	\end{center}
\end{table}

We tested \modelname{} on the publicly available DISC dataset\footnote{\texttt{\url{https://dx.doi.org/10.21227/2gm7-9z72}}}~\cite{pegoraro2022disc}. DISC contains, among other data, $416$ IEEE~802.11ay \ac{cir} sequences at $60$~GHz, obtained at a fixed sampling rate of $T_c = 0.27$~ms using a monostatic \ac{jcs} \ac{sdr} platform. \ac{cir} estimates are obtained in a standard-compliant fashion, using so-called TRN fields of pilot symbols appended as trailers to IEEE~802.11ay packets. The \ac{cir} sequences contain signal reflections on humans performing four different activities: \textit{walking}, \textit{running}, \textit{waving hands} and \textit{sitting down/standing up}.
The dataset includes data from $7$ different subjects. A summary of the parameters of the DISC \ac{cir} data is provided in \tab{tab:disc-params}, while for additional details regarding the data collection and the experimental testbed we refer to~\cite{pegoraro2022disc, pegoraro2023rapid}.
We remark that sequences have different durations, ranging from $0.52$ to $9.22$ seconds, and that the dataset is \textit{unbalanced}, with more samples belonging to \textit{walking} compared to the other activities. 

\subsection{Training details and model parameters}
\label{sec:model_specifics}

\subsubsection{Dataset splitting and preparation} 
\begin{figure}
    \centering
    \includegraphics[width= 0.9 \linewidth]{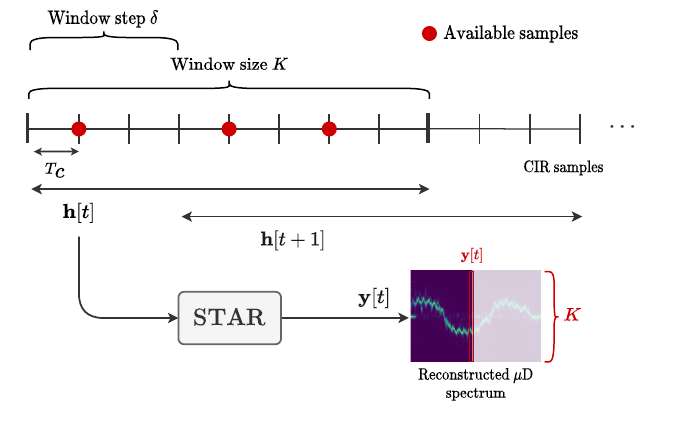}
    \caption{Framing \ac{cir} samples into subsequent windows to be processed by \modelname{}.}
    \label{fig:STAR-input-pipeline}
\end{figure}

We process the \ac{cir} data in windows of $K=64$ steps in the temporal grid with spacing $T_c$, where each new window is shifted to the right by $\delta=32$ samples with respect to the previous one, leading to $K-\delta$ overlapping samples between every two adjacent windows, as illustrated in \fig{fig:STAR-input-pipeline}. Therefore, each of the $416$ sequences in DISC provides tens to hundreds of input \ac{cir} windows, where the exact number depends on the duration of the sequence.  
The whole set of \ac{md} sequences is split into non-overlapping training, validation, and test sets, with ratios $0.8, 0.01, 0.19$. The choice of these ratios is motivated later in this section. We split the dataset at the sequence level, before extracting the \ac{cir} windows,  as: (i)~we want to ensure a sufficient level of diversity between the training and validation/test sets, and (ii)~the training process for \modelname{} is \textit{sequential}, i.e., the \ac{cir} windows must be processed in their original temporal ordering.
In the training and evaluation of \modelname{}, we obtain incomplete \ac{cir} measurement patterns by randomly removing samples from the DISC \ac{cir} sequences.
During training, we provide the network with a wide range of diverse sampling patterns, with different sparsity levels. For this, we dynamically augment our training set by generating random binary masks applied to each \ac{cir} input window. The masks are generated by first sampling a mask probability $p \sim \mathcal{U}(0, 0.9)$, and then setting each element of the mask to $0$ with probability $p$. This augmentation procedure makes the \modelname{} training process extremely robust to overfitting, as the probability that two identical input sequences are presented to the network is negligible. This is the reason why we select a smaller validation set compared to the test set: We give more importance to obtaining a reliable estimate of the final metrics (guaranteed by a larger sample) than to evaluating the validation performance, as we observe no overfitting in our experiments. Moreover, despite being a $0.01$ fraction of the whole dataset, our validation set contains $2837$ \ac{cir} windows. We found this to be a sufficient sample size to tune the hyperparameters of \modelname{} and to monitor its generalization capabilities during training.

Furthermore, to address the unbalanced nature of the dataset, we performed random oversampling of the \ac{md} spectrograms related to the \textit{running} activity. This was done to balance the amount of \textit{walking} and \textit{running} sequences in the training set. This oversampling procedure was crucial: we empirically observed that the dataset being unbalanced, combined with the presence of higher Doppler frequency components in \textit{running} compared to the other activities (due to the higher movement speed of the limbs) led to poor reconstruction of the \textit{running} \ac{md} sequences. 
The oversampling procedure provides an effective solution to this issue, and thanks to the random mask generation there is no repetition of the same input \ac{cir} samples during training, which could cause overfitting.

\subsubsection{Ground truth and loss function} To train the model, we first defined the ground truth as the \ac{md} spectrum reconstructed by the \ac{iht} algorithm at convergence, \textit{when provided with a complete window of \ac{cir} samples}, i.e., when all the measurements are available. This served as reference data to train \modelname{}. We denote the ground truth \ac{md} and the reconstructed \ac{dft} at time step $t$ as $\mathbf{y}_{\rm gt}[t]$ and $\mathbf{z}_{\rm gt}[t]$, respectively. Our model was trained to faithfully reconstruct both the sparse reconstruction and the final \ac{md} spectrum, by minimizing the \ac{mse} between (i)~the ground truth \ac{md}, $\mathbf{y}_{\rm gt}[t]$, and its reconstruction, $\mathbf{y}[t]$, and (ii)~the ground truth \ac{dft} $\mathbf{z}_{\rm gt}[t]$ and $\mathbf{z}[t]$. We denote these two loss terms as $\mathcal{L}_{\mu\mathrm{D}}[t]$ and $\mathcal{L}_{\rm IHT}[t]$, respectively.
The final training loss of our model is then
\begin{equation}
    \mathcal{L}[t] = \alpha \mathcal{L}_{\ac{md}}[t] + \beta \mathcal{L}_{\rm IHT}[t],
\end{equation}
where $\alpha, \beta >0$ are used to tune the relative importance of the two losses.
Experimental results showed that putting more emphasis on the reconstruction of the \ac{md} spectrum ($\mathcal{L}_{\ac{md}}[t]$) yielded better performance, but $\mathcal{L}_{\rm IHT}[t]$ is still useful to provide additional training feedback on the \ac{liht} block, so we set $\alpha=0.9, \beta=0.1$, as this combination of weights led to the best results. 

In the LIHT module, we set the reciprocal of the step size $\mu=20$, while in the attention block we consider $N_p = 6$ past \ac{md} windows.
The model was trained for $5$ epochs, using the Adam optimizer~\cite{kingma2015adam} with a learning rate of $2\cdot 10^{-4}$, on a NVIDIA RTX3080 GPU. In \tab{tab:tab-parameters}, we summarize all the relevant hyperparameters that were used in our experiments. \rev{Note that, in the deep unrolling framework, parameters of the standard \ac{iht} algorithm (e.g., the sparsity parameters $\Omega$ and the inverse \ac{liht} step size $\mu$) can be seen as hyperparameters of the unrolled neural network, which are optimized using a validation set and kept \textit{constant} during training. This distinction is done to set such parameters apart from the neural network weights and learnable parameters, which are instead optimized during training using backpropagation.}

\begin{table} [t!]
\small
	\begin{center}
		\begin{tabular}{lcr}
\toprule
\multicolumn{3}{c}{{\bf \rev{\modelname{}  hyperparameters}}} \\
\midrule
              Window length & $K$ & 64 \\
              Window shift & $\delta$ & 32 \\
              Sparsity parameter & $\Omega$ & 5 \\
              Inverse \ac{liht} step & $\mu$ & $20$\\
              No. of past windows & $N_p$ & 6 \\
              \ac{md} loss weight & $\alpha$ & 0.9 \\
              \ac{iht} loss weight & $\beta$ & 0.1\\
 \bottomrule
\end{tabular}
	\end{center}
	\caption{\modelname{} hyperparameters used in the implementation.} 
	\label{tab:tab-parameters}
\end{table}

\subsection{\ac{md} reconstruction results}
\label{sec:results_quality}

\subsubsection{Reconstruction quality}

As a first evaluation step, we compare the reconstructed \ac{md} spectrum to the ground truth using two different metrics: the \ac{rmse} and the \ac{ssim}~\cite{wang2004image}. Note that in all our results the ground truth \ac{md} spectra have been normalized in the interval $[0,1]$, using min-max normalization as in \cite{pegoraro2023rapid}.
The \ac{rmse} treats all the components of the two signals equally, without considering any specific features of the \ac{md} data. However, this is not always the best way of assessing the quality of reconstruction: When the goal is to evaluate the human-perceived signal quality, the \ac{rmse} has often proven to be inadequate~\cite{wang2009mean}.
Therefore, we also compute the \ac{ssim}, which is based on a combination of several visual aspects like lightness, contrast, and structural information of the reconstructed \ac{md} spectrum, and provides a more reliable measure of perceptual fidelity.

\subsubsection{Comparison with state-of-the-art solutions} \label{sec:SOTA_comparison}
In our numerical analysis, we provide a comparison between our model, \ac{iht}~\cite{pegoraro2022sparcs}, and \ac{dust}~\cite{de2023designing}, which is the most recent approach for sequential sparse recovery based on deep unrolling and attention in the context of video processing. 
\modelname{} and \ac{dust} differ under several aspects. First, \ac{dust} applies the self-attention operation across the whole input frames sequence, thus looking for correlations between the current frame and both past and future inputs. While this approach provides richer modeling of the correlations present in the sequence, it is impractical for real-time \ac{md} reconstruction, since it is \textit{non-causal} and requires knowing the future input frames. 
Therefore, for the sake of a fair comparison, we constrain the attention steps of \ac{dust} to only operate on the same past windows considered by \modelname{}. 

Furthermore, another distinction lies in the domain in which the attention operation is applied. Unlike \modelname{}, which directly applies the attention operation between \ac{md} spectra, \ac{dust} reconstructs the original signal and computes the correlations in the time domain. Specifically, the \ac{dust} attention step is described by the following equation
\begin{align}
    \nonumber \mathbf{z}[t] &= \xi \sum_{i = 1}^{N_p} \frac{ e^{\mathbf{z}[t-i]^T \mathbf{W}^T  \mathbf{W} \mathbf{z}[t]}}{\sum_{u = 1}^{N_p} e^{\mathbf{z}[t-u]^T \mathbf{W}^T  \mathbf{W} \mathbf{z}[t]} } \mathbf{z}[t-i]\\
    &= \xi\mathbf{Z}[t]^T \mathrm{Softmax}\left(\mathbf{Z}[t]^T \mathbf{W}^T  \mathbf{W} \mathbf{z}[t] \right),
\end{align}
where $\mathbf{Wz}[t]$ is the reconstructed signal in the time domain, \mbox{$\mathbf{Z}[t] = \left[ \mathbf{z}[t-1], \dots, \mathbf{z}[t-N_p]\right]^T$}, and $\xi$ is a trainable parameter.
In the original version of \ac{dust}, the output of the attention layer is used to initialize a \ac{lista} layer, which is then trained to solve the sparse reconstruction problem. We underline that this is a key distinction between \ac{dust} and \modelname{}, as in the former the attention is applied at the input section, whereas in the latter it is employed at the end of the processing chain. Now, to understand the impact of this architectural choice, for \ac{dust}, we replace the \ac{lista} module with a \ac{liht} module. In this case, \ac{dust} and \modelname{} reconstruct the micro-Doppler spectrum using a similar approach and their main difference resides in the positioning of the attention mechanism. Moreover, we also consider a \ac{dust} variant, which we denote by \ac{dust}-V2, which still uses the attention at the input, but the corresponding operations are executed in the frequency domain (as we do in \modelname{}).

The experimental results show almost identical performance for the two \ac{dust} variants, which proves that the gain brought by \modelname{} is not due to the domain (time versus frequency) in which the attention is performed. It is also important to observe that \ac{dust} requires \textit{at least two} iterations of \ac{liht} to be performed to successfully compute its attention weights (see Section~3.1 of \cite{de2023designing}).
This has to be taken into account when comparing it to \modelname{}, which instead entails a single \ac{liht} iteration. 

\rev{
To provide a broader benchmark of \modelname{}'s performance, we also compare it to established compressed sensing algorithms, namely \ac{omp} and \ac{lasso} (or $\ell 1$-norm minimization). Note that we do not evaluate \ac{omp} and \ac{lasso} against \ac{iht}'s ground truth, as for \modelname{} and \ac{dust}, but against their own ground truth reconstructions obtained with a full measurement window. This choice is motivated by the fact that, unlike \modelname{} and \ac{dust} which are based on \ac{iht}, \ac{omp} and \ac{lasso} may yield intrinsically different reconstructions compared to \ac{iht} since they are based on different optimization approaches. Our comparison method removes this mismatch and only considers the degradation due to the fewer \ac{cir} samples.  
} 

\begin{figure}[t!]
    \centering
    \includegraphics[width=\linewidth]{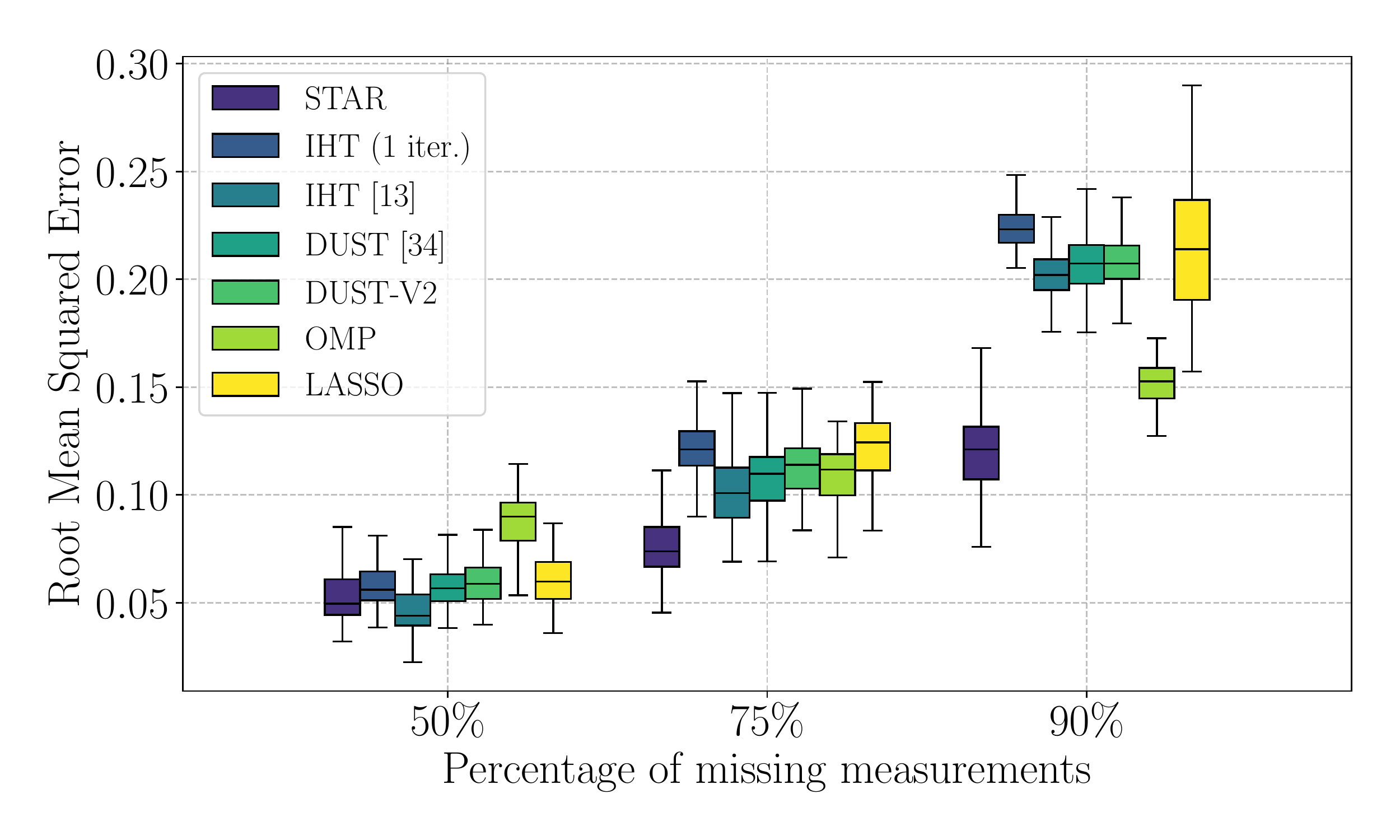}
    \caption{\rev{Evolution of the \ac{rmse} over the test set for increasing percentages of missing measurements, achieved by \modelname{}, in comparison with \ac{iht} at convergence, \ac{iht} stopped after one iteration, DUST, DUST-V2, OMP, and LASSO algorithms.}}
    \label{fig:plot-rmse}
\end{figure}

\begin{figure}[t!]
    \centering
    \includegraphics[width=\linewidth]{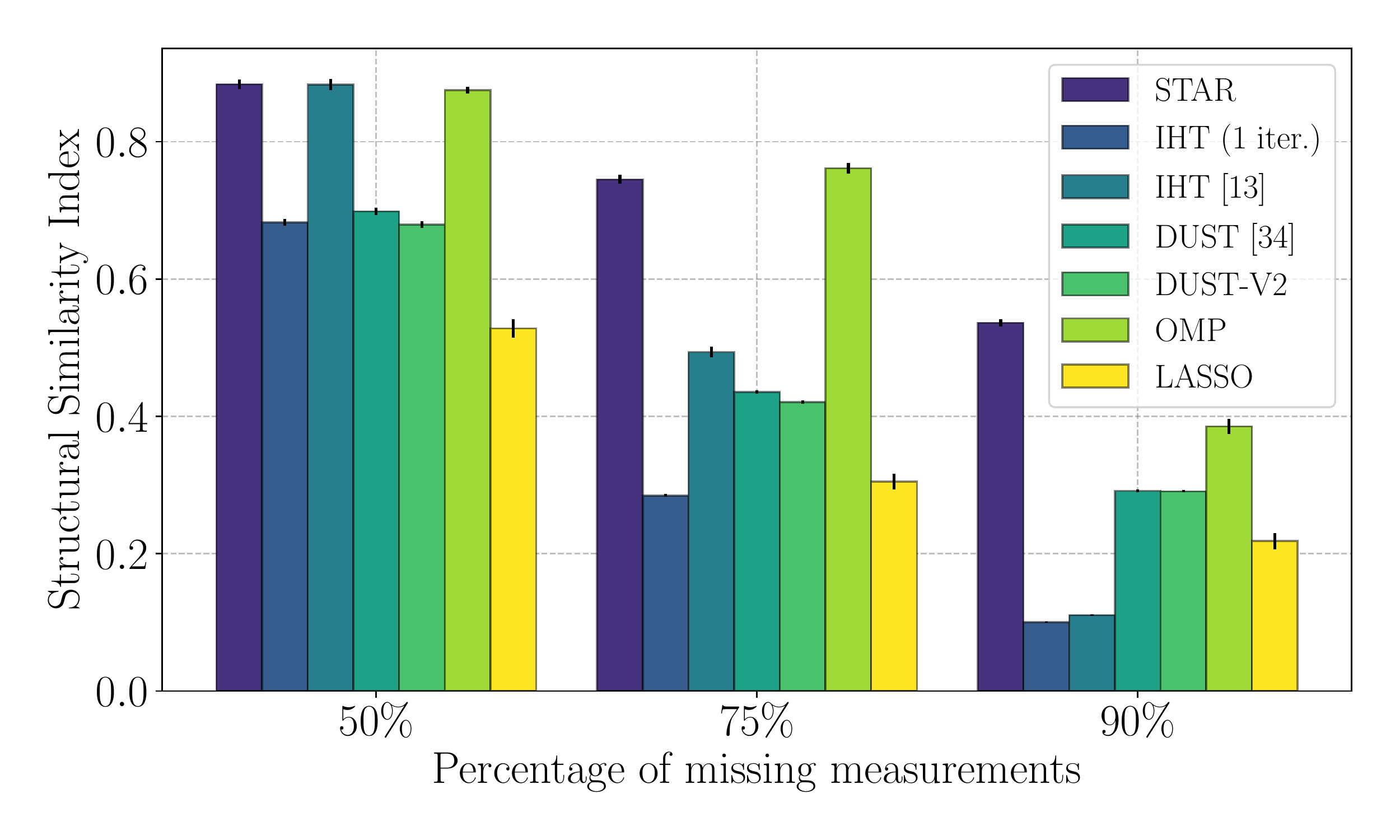}
    
    \caption{Evolution of the \ac{ssim} over the test set for increasing percentages of missing measurements, achieved by \modelname{}, in comparison with \ac{iht} at convergence, \ac{iht} stopped after one iteration, DUST, DUST-V2, OMP, and LASSO algorithms.}
    \label{fig:plot-ssim}
\end{figure}
\begin{figure*}[t]
	\begin{center}   
            \includegraphics[width=\linewidth]{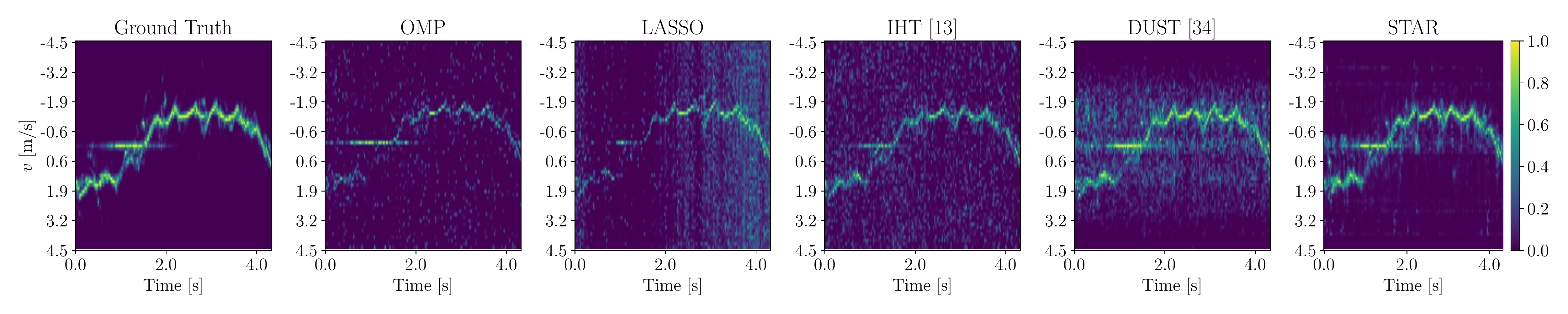}
		\caption{\rev{Comparison of different sparse reconstructions provided by \ac{omp}, \ac{lasso}, \ac{iht}, \ac{dust} and \modelname{}, when provided with \ac{cir} windows with $90\%$ missing channel measurements.}}
		\label{fig:md-comparison}
	\end{center}
	\vspace{-0.5cm}
\end{figure*}
\fig{fig:plot-rmse} and \fig{fig:plot-ssim} show the \ac{md} reconstruction errors in terms of \ac{rmse} and \ac{ssim}, respectively. 
\rev{Error bars in \fig{fig:plot-ssim} represent the standard error dispersion measure, i.e., $\sigma/\sqrt{n_{\rm test}}$, where $\sigma$ and $n_{\rm test}$ are the standard deviation and number of samples in our test set, respectively. }

These results show that \modelname{} performs close to \ac{iht} when the channel is densely sampled (low sparsity), which is the best-case scenario for \ac{iht}, and where \modelname{} output gets very close to the ground truth reconstructions. However, as the number of available samples decreases, \modelname{} consistently outperforms \ac{iht} stopped after one iteration (\ac{iht} $1$ iter.), and \ac{iht} at convergence, both in terms of \ac{rmse} and \ac{ssim}. 
\modelname{} also steadily outperforms both \ac{dust} variants.
\rev{To visually compare the quality of the reconstructions for a highly incomplete window ($90\%$ missing measurement), in \fig{fig:md-comparison} we show the \ac{md} spectrograms reconstructed using \modelname{}, \ac{iht} at convergence, \ac{dust}, \ac{omp} and \ac{lasso}.}

Two key observations are in order.

\noindent\textit{(i) Placement of the attention layer}: In the DL literature, attention layers are typically placed at the very beginning of the processing chain. For example, the classical transformer architecture applies multiple self-attention layers in its input encoder module to learn contextualized features that are exploited by the subsequent layers. Similarly, DUST applies the self-attention mechanism at the input, to provide good initialization for the following \ac{lista} algorithm. This is the key difference that sets \modelname{} apart from \ac{dust}: Our model is architecturally different, as the attention mechanism is utilized towards the output of the processing chain rather than at its input, to refine the coarse-grained reconstruction obtained by the first unrolled layer. This choice was found to be highly effective for the reconstruction of highly sparse \ac{md} spectra.

\noindent\textit{(ii) Role of the unrolled layer}: Existing approaches primarily rely on the predictions from the unrolled algorithms. This is not the case for \modelname{}, where the primary role of the unrolled \ac{liht} layer is to provide a first approximation of the solution. For the considered sensing application, experimental results revealed that just providing a good initialization to the \ac{liht} layer (i.e., using attention at its input) is ineffective, and leads to poor-quality solutions.

\subsubsection{Improved activity recognition capabilities}\label{sec:impr-act-rec}

In this section, we evaluate the quality of the reconstructed \ac{md} signatures by using them as the input to a standard \ac{jcs} application that performs human activity recognition.
In the literature, this is typically implemented by applying a \ac{cnn} classifier on the \ac{md} spectrograms~\cite{vandersmissen2018indoor, pegoraro2023rapid, pegoraro2022sparcs}.
We show that \modelname{} provides a substantial improvement in the activity recognition task when only a few channel measurements are available. To this end, the following methodology was adopted:
\begin{enumerate}
    \item[i.] We built a set of training data pairs $\mathcal{X}_{\text{train}} = \left\{(\mathbf{X}_i,\ell_i)\right\}_{i=1}^{N_{\rm train}}$. Each $\mathbf{X}_i$ is a crop of a ground truth \ac{md} spectrogram, obtained by applying \ac{iht} until convergence on a sequence of $\Gamma = 200$ consecutive \ac{cir} windows. As a result, all the spectrogram crops in $\mathcal{X}_{\text{train}}$ have a time duration of $((\Gamma - 1) \delta + K)  T_c \approx 1.7$ seconds of measurements, see \fig{fig:STAR-input-pipeline}.
   $\ell_i$ is the activity label of the corresponding crop $\mathbf{X}_i$, which can be one among \textit{walking}, \textit{running}, \textit{waving hands} or \textit{sitting/standing up}. 
    \item[ii.] We built a set of test data pairs $\hat{\mathcal{X}}(p) = \{(\hat{\mathbf{X}}_i,\hat{\ell}_i)\}_{i=1}^{N_{\rm test}}$, where each $\hat{\mathbf{X}}_i$ is a crop of $\Gamma$ consecutive \ac{md} windows reconstructed by our model from an incomplete sequence of \ac{cir} samples, with a level of sparsity $p$. We stress that this set is built using the \textit{test} \ac{cir} sequences, hence there is no overlap between the $\hat{\mathbf{X}}_i$ and the training set of \modelname{}.
    \item[iii.] We then trained a \ac{cnn} classifier on the task of activity recognition on set $\mathcal{X}_{\text{train}}$. The structure of this neural network classifier is summarized in \tab{tab:cnn-architecture}. We trained this \ac{cnn} for $100$ iterations, using the Adam optimizer with a learning rate of $10^{-3}$. 
    \item[iv.] Finally, we obtained the \ac{cnn} output classifications on the test set $\hat{\mathcal{X}}(p)$, and computed the global and per-class F$1$-scores as the evaluation metrics.
\end{enumerate}
    \begin{table} [t!]
\small
	\begin{center}
		\begin{tabular}{lcc}
\toprule
\textbf{ Layer} &\textbf{In Channels} &  \textbf{Out Channels}  \\
\toprule
    \texttt{Conv_1 (\ac{relu})} & $1$ &$8$  \\
    \texttt{Conv_2 (\ac{relu})} & $8$ &$16$  \\
    \texttt{Conv_3 (\ac{relu})} & $16$ &$32$  \\
    \texttt{Conv_4 (\ac{relu})} & $32$ &$64$  \\
    \texttt{Conv_5 (\ac{relu})} & $64$ &$128$  \\
    \texttt{Conv_6 (\ac{relu})} & $128$ &$128$  \\
    \texttt{Flatten} & - & -  \\
    \texttt{Linear (\ac{relu})} & $512$& $64$ \\
    \texttt{Dropout (p=0.2)} & - & -  \\
    \texttt{Linear}& $64$ & $4$  \\
    
 \bottomrule

\end{tabular}
	\end{center}
	\caption{Baseline CNN Architecture used in the evaluation. All convolutional layers use $3 \times 3$ kernels with a stride of $2$.} 
	\label{tab:cnn-architecture}
\end{table}

\begin{figure}[t!]
    \centering
    \includegraphics[width=\linewidth, trim=0cm 0.5cm 0cm 2.0cm, clip]{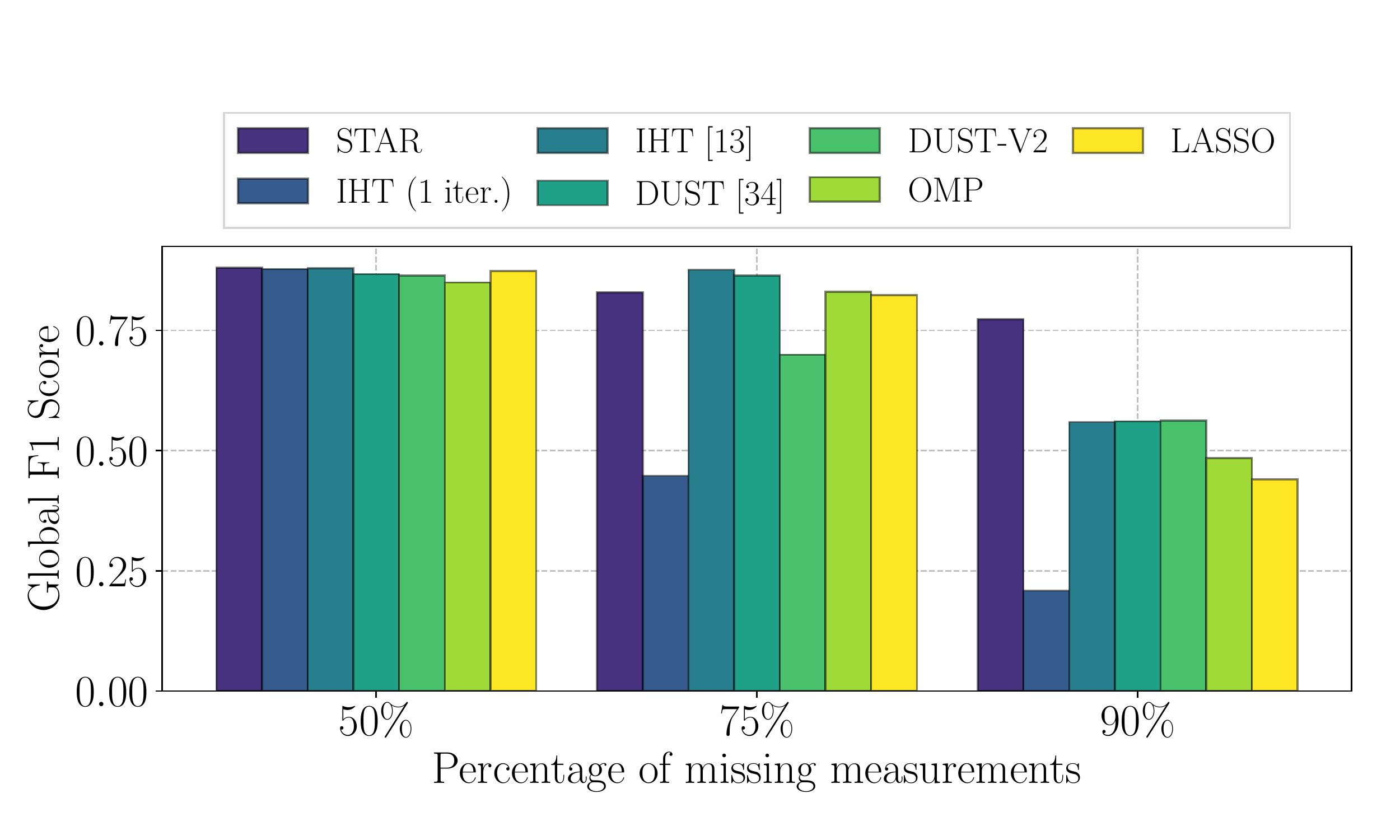}
    \caption{Global F$1$-score for different sparsity levels. }
    \label{fig:cnn-f1-global}
\end{figure}

\begin{figure}[t!]
    \centering
    \includegraphics[width=\linewidth, trim=0cm 0.5cm 0cm 1.5cm, clip]{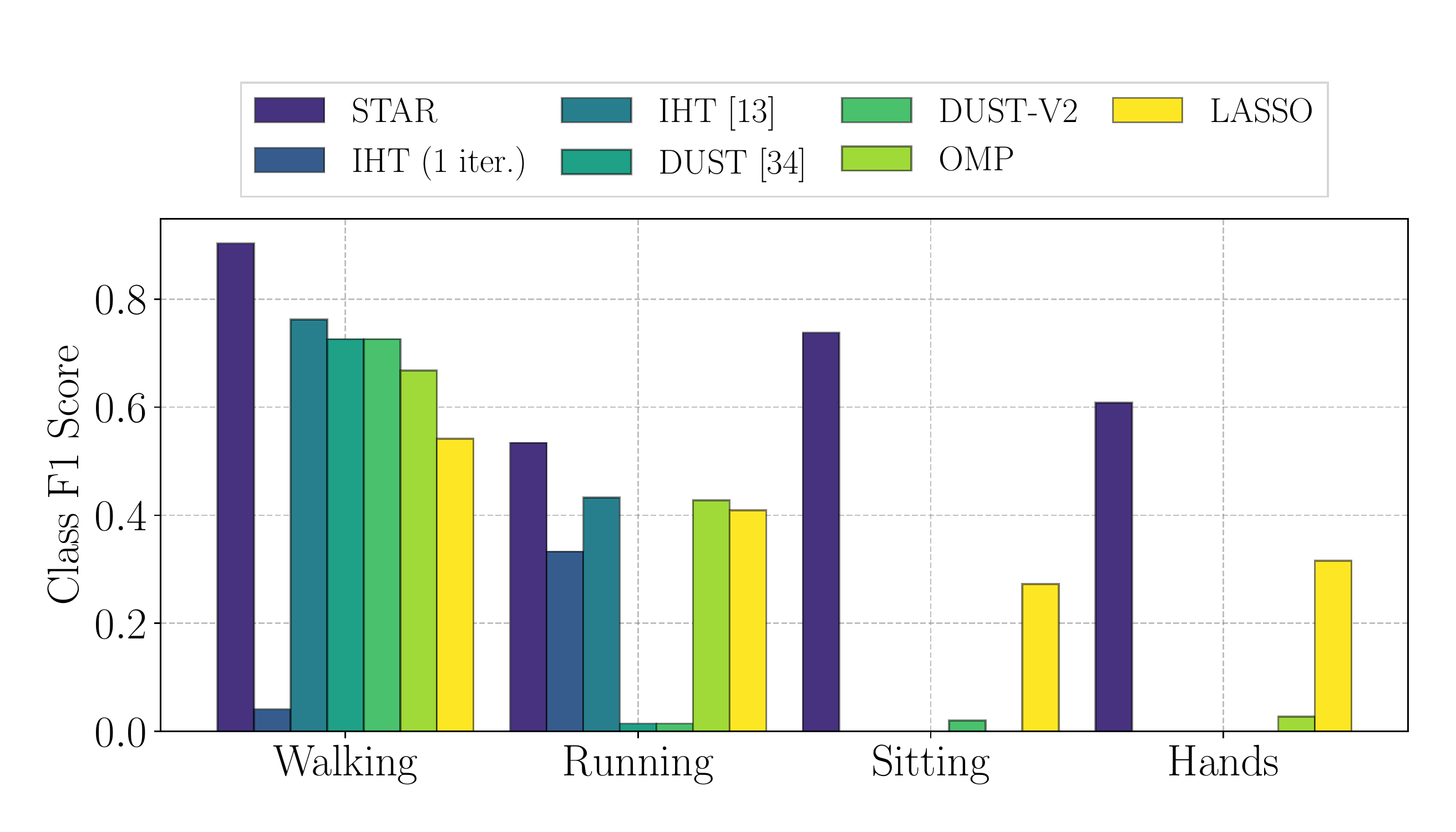}
    \caption{Per-class F$1$-score with $90\%$ incomplete channel measurements.}
    \label{fig:cnn-f1-classes}
\end{figure}

\begin{figure*}[t]
	\begin{center}   
            \includegraphics[width=\linewidth]{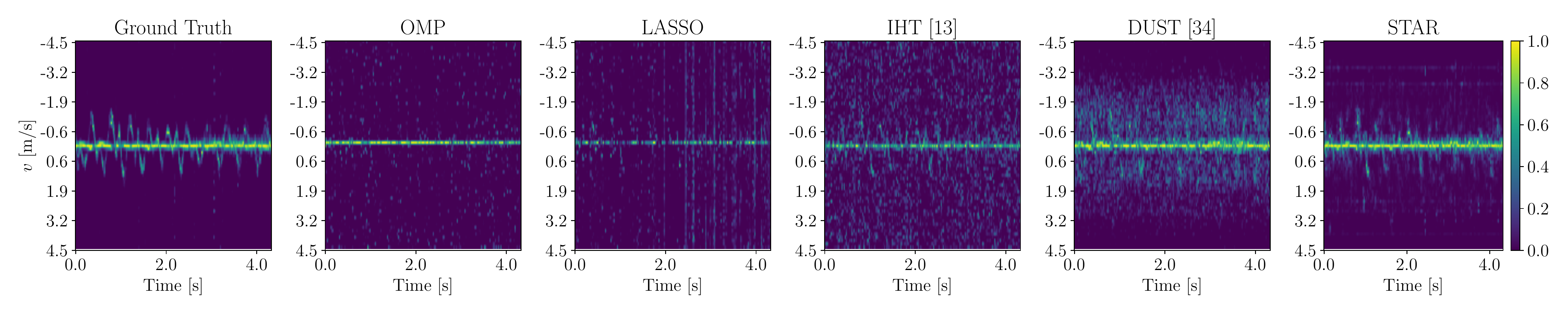}
		\caption{Comparison of different sparse reconstructions for the \textit{Waving hands} activity, with $90\%$ missing input channel measurements. \modelname{} is capable of retaining the characteristic shape of the activity while removing the majority of the background noise.}
		\label{fig:md-comparison-hands}
	\end{center}
\end{figure*}

In \fig{fig:cnn-f1-global} and \fig{fig:cnn-f1-classes} we report the F$1$-scores achieved by \modelname{}, \ac{iht} at convergence, \ac{iht} stopped after one iteration, \ac{dust}, and \ac{dust}-V2.
In terms of global F$1$-score, results show that our method performs similarly to \ac{iht} and \ac{dust} at sparsity levels of $50$\% and $75$\%, but it greatly outperforms them with a $90\%$ measurement sparsity. 
The comparable performance when many measurements are available is due to the robustness of \acp{cnn} to the presence of noise in the input data. Indeed, the reconstructions of all the evaluated models with low sparsity are only slightly noisier than those obtained from a full measurement window. 
However, at extreme sparsity levels, the quality of the reconstructions provided by the other models drops significantly. Conversely, the reconstructions provided by \modelname{} preserve the specific features of human movements, reaching an F$1$-score close to $0.8$. 
\rev{A deeper insight into this can be gained by observing the values of the class-specific F$1$-scores (see \fig{fig:cnn-f1-classes}). The class-specific F$1$-score evaluates the capability of the classifier to recognize each class individually, whereas the global F$1$-score is a measure of the aggregate performance. The class-specific F$1$-score is useful to understand which classes are more affected by the high number of missing measurements. With $90$\% missing measurements, \modelname{} largely outperforms its competitors on each activity. The gain is especially evident for those activities involving fine-grained movements, i.e., \textit{sitting/standing up} and \textit{waving hands}. \ac{dust}, \ac{iht}, and \ac{omp} instead fail to provide discriminative reconstructions, obtaining an F$1$-score of $0$. This is due to the insufficient reconstruction quality: As shown in \fig{fig:md-comparison-hands}, in \ac{md} recovered by \ac{dust}, \ac{iht}, and \ac{omp} the movement features are hidden by severe reconstruction artifacts. Conversely, \ac{lasso} has lower (not exceeding $55$\%) but more consistent F$1$-scores across activities.

\modelname{} achieves good performance with over $0.6$ F$1$-score for all classes. This shows that the gain brought by \modelname{} is much more significant than the $0.2$ gain shown in \fig{fig:cnn-f1-global}. \modelname{} is the only model that can recover the \ac{md} features of small movements, which is a key enabler for gesture recognition applications. \fig{fig:md-comparison-hands} provides a qualitative comparison of \ac{md} reconstructions relative to the \textit{waving hands} activity.} 

\subsection{Ablation studies}
\label{sec:ablation}
We perform an ablation study by evaluating several variations of our model, obtained by modifying or removing different components of the architecture, to assess their impact on the final performance. 
The considered variations are: 
\begin{itemize}
    \item \modelname{} with $N_p = 1, 3, 9$; 
    \item \modelname{} where we replace matrix $\mathbf{I} - \mathbf{W}^T\mathbf{W}$ in \eq{eq:liht-module} with matrix $\mathbf{S}\in \mathbb{R}^{2K \times 2K}$, containing $4K^2$ additional learnable weights. We refer to this variant as ``Learn $\mathbf{S}$".
    \item a single \ac{liht} block, i.e., \modelname{} without the attention and solution refinement blocks. We call this variation ``No Attention";
    \item \modelname{} where the solution refinement block is modified by removing the multiplicative branch, only retaining the additive modification of the \ac{liht} solution (``Only Add");
\end{itemize}

In \tab{tab:tab-ablation-all}, we report the \ac{rmse} and \ac{ssim} metrics, as well as the global F$1$-score achieved by the \ac{cnn} classifier applied to the reconstructed \ac{md} signatures, for each model variation. 

Our results reveal three main insights regarding our model. Firstly, increasing $N_p$ above $6$ does not yield a consistent improvement for all metrics. Indeed, the best \ac{cnn} F$1$-score with $90$\% sparsity, as well as the best \ac{rmse} values are obtained with $N_p=6$. Moreover, larger values of $N_p$ require storing a longer sequence of past outputs and increase the complexity of the attention mechanism. This is not necessarily beneficial, since the autocorrelation of \ac{md} spectra quickly goes to zero as the time lag increases. Therefore, we choose $N_p=6$ as the default configuration for \modelname{}. 

Secondly, learning matrix $\mathbf{S}$ provides a performance improvement in terms of \ac{ssim} and \ac{cnn} F$1$-score at lower sparsity levels, but it is not beneficial with a sparsity of $90$\%.
The additional cost of learning $4K^2$ additional weights, and the fact that learning $\mathbf{S}$ makes the model less interpretable has to be considered as well. Hence, in \modelname{} we decided not to learn $\mathbf{S}$, and to use matrix $\mathbf{I} - \mathbf{W}^T\mathbf{W}$ in \eq{eq:liht-module} instead.

Finally, the results for ``No Attention" and ``Only Add" show that the proposed additive and multiplicative refinement block, based on a context vector provided by attention, brings a substantial improvement. This is especially evident for high percentages of missing input measurements. \rev{The reason for this improvement lies in the additional information provided by attention, which is particularly important in the presence of highly incomplete input measurements.}  

\begin{table*} 
\small
	\begin{center}
		\begin{tabular}{ccccccccccc}
\toprule
& \multicolumn{3}{c}{{\bf \ac{rmse}}} & \multicolumn{3}{c}{{\bf \ac{ssim}}}  & \multicolumn{3}{c}{{\bf CNN F$1$-score}} & \multirow{2}{*}{\textbf{No. parameters}} \\
\cmidrule(lr){2-4} \cmidrule(lr){5-7} \cmidrule(lr){8-10}
 \textbf{Missing measurements [\%]} & $50\%$ & $75\%$ & $90\%$ & $50\%$ & $75\%$ & $90\%$ & $50\%$ & $75\%$ & $90\%$ \\
\midrule
              $N_p=1$& 0.0564 &   0.0828 & 0.1379 &  0.806 & 0.608 & 0.374 & 0.875 &
  0.814 & 0.764 & 24,640  \\
              $N_p=3$ &0.0546 & 0.0784 & 0.1252 & 0.859 & 0.687 & 0.433 & 0.873 &
  0.831 &  0.766 & 24,640 \\
   $N_p=9$ &0.0548 & 0.0784 & 0.1214 & 0.887 & 0.760 & \textbf{0.581} & \textbf{0.882} &  0.829 &  0.757 & 24,640  \\
              Learn $\mathbf{S}$  &0.0578 &             0.0846 &              0.1412 & \textbf{0.897} & \textbf{0.771} & 0.488  & 0.876 &
  \textbf{0.859} &  0.330 & 41,024  \\
              No Attention&0.0635 &             0.1128  &              0.1999 & 0.724 & 0.492 & 0.319 & 0.845 & 0.701 & 0.582 & 16,384  \\
              Only Add&0.0636  &             0.1128   &              0.1998 & 0.714 & 0.479 & 0.230 & 0.841 &  0.700 &  0.586 & 20,544 \\
              \modelname{} & \textbf{0.0545} & \textbf{0.0779} & \textbf{0.1213} & 0.884 & 0.745 & 0.536 & 0.881 & 0.829 & \textbf{0.773} & 24,640 \\
 \bottomrule
\end{tabular}
	\end{center}
	\caption{Results of the ablation studies in which we compare different variations of \modelname{}. We highlight the best values using a \textbf{bold} font.} 
	\label{tab:tab-ablation-all}
\end{table*}

\rev{

\subsection{Test in a different environment}\label{sec:diff-env} 
To assess the ability of our model to generalize across different environments, we evaluate \modelname{} on \ac{cir} sequences collected in a different room (called \textit{test room}) than the one it was trained on (\textit{training room}). The data from the test room is also provided in the DISC dataset\cite{pegoraro2022disc} and contains measurements of a single human subject performing the same set of actions of the training set. Following the same methodology of \secref{sec:results_quality}, we evaluate the model in terms of \ac{md} spectrogram reconstruction error and F$1$-score on the activity recognition task using the \ac{cnn} of \tab{tab:cnn-architecture}. In \tab{tab:star-generalization-results}, we compare the values of \ac{rmse}, \ac{ssim}, and \ac{cnn} F$1$-scores obtained on reconstructions coming from the training room and the test room. 
The results are in line with those presented in \secref{sec:results_quality}, showing the same reconstruction quality, and similar degradation on the classification task when providing the \ac{cnn} model with reconstructions from $90$\% missing measurements. This demonstrates that \modelname{} generalizes beyond the specific environment used during training.

\begin{table}[t!] 
\footnotesize
    \caption{\rev{\modelname{} performance in a different environment. We report \ac{md} reconstruction quality metrics (\ac{rmse} and \ac{ssim}) in the training room and the (unseen) test room.}} \label{tab:star-generalization-results}
    \vspace{-0.25cm}
	\begin{center}
		\begin{tabular}{ccc}
			\toprule	
			 & \textbf{RMSE} & \textbf{SSIM} \\
            \cmidrule(lr){2-3}
			\textbf{Training room} & $0.121 \pm 0.002$&$0.536 \pm 0.005$\\
                \textbf{Test room} & $0.115 \pm 0.004$&$0.479 \pm 0.006$\\
			\bottomrule
		\end{tabular} 	
 
	\end{center}
\end{table}

\begin{table}[t!] 
\footnotesize
    \caption{\rev{CNN F$1$-score on ground truth (\ac{iht} full) and STAR reconstructions obtained in the test room.}} \label{tab:star-generalization-results2}
    \vspace{-0.25cm}
	\begin{center}

        \begin{tabular}{cccccc}
			\toprule	
			 & \textbf{Walking} & \textbf{Running} & \textbf{Sitting} & \textbf{Hands} & \textbf{Global} \\
            \cmidrule(lr){2-6}
			 \textbf{IHT (full)} &$0.896$ & $0.902$ & $0.607$ &$0.588$&$0.750$ \\
                \textbf{STAR (90\%)} & $0.829$ & $0.750$ & $0.557$ &$0.432$&0.652 \\
			\bottomrule
		\end{tabular} 
	\end{center}
\end{table}

\subsection{Computational complexity and model size}\label{sec:comp-compl-size}

In this section, we show that \modelname{} is a lightweight model both in terms of computational complexity and model size (intended as the number of learnable parameters).  

\subsubsection{Computational complexity} The complexity of \modelname{} is in the same order of just one \ac{iht} iteration, making it well suited for real-time applications.
We start by analyzing the complexity of \ac{iht}. From \eq{eq:liht-module}, one can see that the number of operations required at each iteration is asymptotically dominated by the matrix product $\mathbf{W}^T\mathbf{W}$. As $\mathbf{W} \in \mathbb{C}^{M_t \times K}$, the number of operations involved in a single iteration is in the order of $K^2 M_t$. Overall, considering $N$ \ac{iht} iterations, the complexity is $\mathcal{O}(N K^2M_t)$.  

We can now analyze the complexity of \modelname{}, by deriving the number of computations for each module:
\begin{enumerate}
    \item The \ac{liht} block has the same complexity as a single \ac{iht} iteration, namely, $\mathcal{O}(K^2 M_t)$.
    \item The attention mechanism performs $N_p$ dot products between vectors of dimension $K$, resulting in a complexity of $\mathcal{O}(KN_p)$. Linearly combining  the resulting correlations does not increase the complexity order.
    \item The computational complexity of the solution refinement module is determined by the vector-matrix multiplications performed by the two feedforward layers (\eq{eq:sol-refinement}). Recalling that matrices $\mathbf{U},\mathbf{V} \in \mathbb{R}^{2K\times 2K}$, and $\mathbf{a}[t]\in\mathbb{R}^{K}$, the complexity is $\mathcal{O}(K^2)$. 
\end{enumerate}
Combining the above steps we obtain a total complexity of $\mathcal{O}(K^2M_t)$, which represents an $N$-fold gain with respect to applying \ac{iht} to convergence. In practice, this provides a huge speedup in reconstructing the \ac{md} sequences, as \ac{iht} easily takes $N \approx 15$ to $20$ iterations to converge.

A similar analysis can be performed for \ac{dust}, yielding a complexity of $\mathcal{O}(NK^2M_t)$, which is identical to that of \ac{iht}. However, \ac{dust} takes much fewer iterations to provide acceptable results, although a minimum of $2$ is required, as explained in \secref{sec:results_quality}. This shows that \ac{dust} is not faster than our model, even in the best case with $N=2$.

\subsubsection{Model size} 

Our approach is also lightweight in terms of number of learnable parameters. In the \ac{liht} module, which performs the computations described in \eq{eq:liht-module}, the only learnable parameters are the entries of matrix $\mathbf{W}$, resulting in \mbox{$4K^{2} = 16384 $} parameters. As discussed in \secref{sec:ablation}, learning additional weights in this module does not significantly enhance the performance. The attention module does not contain learnable parameters. Finally, the solution refinement block in \eq{eq:sol-refinement} learns the weight matrices $\mathbf{U}$, $\mathbf{V}$, and the bias vector $\mathbf{b}$, resulting in a total of \mbox{$2K^2 + K = 8256$} parameters. Hence, \modelname{} consists of a total of $24640$ learnable parameters ($32$~bit floating point), amounting to $98$~kB memory space.
As a comparison, \ac{dust} has $32769$ parameters, as it learns an additional weight matrix in its unrolling layer. 
}

\begin{figure}[t!]
    \centering
    \includegraphics[width=\linewidth]{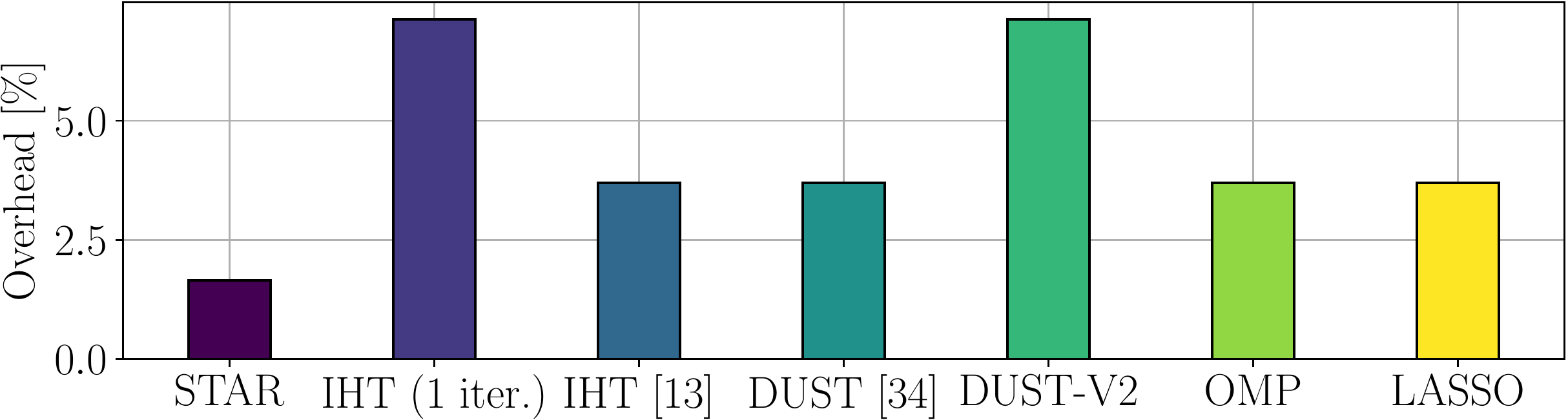}
    \caption{\rev{Overhead on communication introduced by each model to obtain a global F$1$-score over $0.75$. }}
    \label{fig:oh}
\end{figure}

\rev{\subsection{Communication overhead reduction}\label{sec:oh}
Next, we evaluate the impact of applying \modelname{} to a \ac{jcs} system to reduce the overhead introduced by the sensing process on communication. To this end, we consider an IEEE~802.11ay system as a reference. We consider the system parameters of \tab{tab:disc-params} along with processing windows of \mbox{$W=64$} time slots, in which communication payloads with a \ac{psdu} size of $4$~kB are transmitted in standard-compliant packets using \ac{mcs} $9$ (similar but rescaled results are obtained for other \acp{mcs}). Each packet includes preamble fields containing $4352$ $\pi/2$-\ac{bpsk}-modulated symbols~\cite{802.11ay}.
To perform sensing, we follow~\cite{blandino2022tools, pegoraro2023rapid}, assuming that a TRN field with $768$~symbols is appended to the packets. 

For each method, we derive the number of \ac{cir} samples needed to achieve a global F$1$-score threshold of at least $0.75$ on the activity recognition task, by inspecting \fig{fig:cnn-f1-global}. As an example, \modelname{} can reach this threshold even with just $10\%$ available measurements per window, hence it requires $7$ \ac{cir} samples with \mbox{$W=64$}. Conversely, DUST requires $16$ \ac{cir} samples ($25\%$ of the window). We assume that $10$ communication packets are transmitted in the window, serving as the reference number of information bits transmitted by the user.   
The overhead is obtained as the ratio between the length of the \textit{added} PHY layer symbols in the TRN units (for sensing purposes) and the total number of symbols in the packet, considering the number of information bits plus PHY and MAC headers~\cite{802.11ay}. 

\fig{fig:oh} shows the overhead introduced by each \ac{md} reconstruction method based on how many measurements per window it needs to reach the F$1$-score threshold ($0.75$). \modelname{} adds less than half the overhead of the other methods, thus demonstrating its usefulness in lowering the impact of sensing on the communication performance. Note that the specific overhead value depends on the number of communication packets that are to be transmitted in a processing window. Hence, \fig{fig:oh} is intended to provide a comparison among the different methods rather than actual overhead estimates.

As a final note, we remark that, when communication traffic is extremely irregular, \ac{md} reconstruction may require the transmission of \textit{additional} channel estimation fields for sensing, as highlighted in~\cite{pegoraro2022sparcs}. Therefore, it should be considered that, besides a larger overhead, existing methods also entail higher channel occupation, which may cause a communication rate reduction for \textit{other} network nodes in the same radio cell. 
}

\vspace{-0.9\baselineskip}
\section{Concluding Remarks}
\label{sec:conclusion}

In this paper, we tackled the problem of reconstructing \ac{md} spectrograms of human movement from \textit{highly incomplete} channel estimates in a \ac{jcs} system. To this end, we designed and evaluated \modelname{}, an interpretable \ac{nn} architecture that effectively combines deep unrolling of a single iteration of thresholding-based compressed sensing and an attention mechanism that exploits the temporal sequentiality of the \ac{md}. 
The key insight behind \modelname{} design is to use attention to directly enhance the reconstructed spectrum, thus boosting the model's robustness to highly incomplete input measurements. 
Differently from existing standard or learning-based approaches, \modelname{} provides accurate \ac{md} reconstructions even with $90$\%-incomplete channel estimates while retaining a low computational complexity.

\modelname{} is thoroughly evaluated on the publicly available DISC dataset~\cite{pegoraro2022disc}, containing standard-compliant $60$~GHz IEEE~802.11ay \ac{cir} estimates from a physical environment where signal reflections are affected by moving people.
\modelname{} significantly outperforms existing algorithms from the literature in terms of \ac{rmse} and \ac{ssim}. Moreover, when using the reconstructed \ac{md} to perform human activity recognition, state-of-the-art approaches completely fail when only $10$\% of the channel measurements are available ($0$ to $0.3$ F$1$-score), while \modelname{} yields F$1$-scores from $0.5$ to $0.8$.

\rev{Future work includes exploring more advanced ways of integrating signal processing domain knowledge into the \ac{nn} architecture. A possible research direction is to move beyond the deep unrolling paradigm to develop model-based and physics-informed neural networks, which naturally embed equations regulating signal propagation and human movement models into their structure.}

\vspace{-\baselineskip}
\bibliography{references}
\bibliographystyle{ieeetr}

\end{document}